\documentclass{aa}  

\usepackage{graphicx}
\usepackage{txfonts}
\usepackage{hyperref}
\newcommand\Tstrut{\rule{0pt}{2.4ex}}
\newcommand\Bstrut{\rule[-1.ex]{0pt}{0pt}}

\begin{document}

   \title{Weighing protoplanetary discs with kinematics: physical model, method and benchmark}


   \author{B. Veronesi
          \inst{1}
          , C. Longarini\inst{2,3}, G. Lodato\inst{2}, G. Laibe\inst{1}, C. Hall\inst{4}, S. Facchini\inst{2}
          \and
          L. Testi\inst{5,6}
          }

   \institute{Univ Lyon, Univ Lyon1, Ens de Lyon, CNRS, Centre de Recherche Astrophysique de Lyon UMR5574, F-69230, Saint-Genis,-Laval, France\\
              \email{benedetta1.veronesi@gmail.com},
         \and
             Dipartimento di Fisica, Università degli Studi di Milano, Via Celoria 16, 20133 Milano, Italy
        \and
            Institute of Astronomy, University of Cambridge, Madingley Rd, CB30HA, Cambridge, UK\\
             \email{cl2000@cam.ac.uk},
        \and
            Center for Simulational Physics, The University of Georgia, Athens, GA 30602, USA
        \and
            Dipartimento di Fisica e Astronomia “Augusto Righi'' Viale Berti Pichat 6/2, Bologna
        \and 
            INAF, Osservatorio Astrofisico di Arcetri, Largo E. Fermi 5, I-50125, Firenze, Italy\\
             }

   \date{Received XXX; accepted XXX}

 
  \abstract
   {The mass of protoplanetary discs sets the amount of material available for planet formation, determines the level of coupling between gas and dust, and possibly sets gravitational instabilities. Measuring mass of discs is challenging, since it is not possible to directly detect H$_{2}$, and CO-based estimates remain poorly constrained. }
   {An alternative method that does not rely on tracers-to-H$_{2}$ ratios has recently been proposed 
to dynamically measure the disc mass altogether with the star mass and the disc critical radius by looking at deviations from Keplerian rotation induced by the self-gravity of the disc. So far, this method has been applied to weigh three protoplanetary discs: Elias 2-27, IM Lup and GM Aurigae.}
   {We provide here a numerical benchmark of the method by simulating isothermal self-gravitating discs with a range of masses from 0.01 to $0.2 \,M_{\odot}$ with the \textsc{phantom} code and post-process them with radiative transfer (\textsc{mcfost}) to obtain synthetic observations.}
   {We find that dynamical weighing allows to retrieve the expected value of disc masses as long as the disc-to-star mass ratio is larger than $M_d/M_\star=0.05$. The estimated uncertainty for the disc mass measurement is $\sim 25\%$.}
   {}


\titlerunning{Weighing protoplanetary discs with kinematics}
\authorrunning{B. Veronesi et al.}
   \maketitle
%

\section{Introduction}

   A fundamental but still poorly constrained property of protoplanetary discs is their mass. This parameter is of paramount importance because it affects the planet formation process in many different ways. First of all, it determines the amount of material available for the formation of planets \citep{manara18,mulders21}, and, once formed, their migration efficiency (e.g., \citealt{nelson18}). Then, it sets the interaction between gas and dust particles through drag, deeply influencing dust dynamics, its radial drift and its growth (e.g., \citealt{birnstiel10a,birnstiel10b,laibe14,gonzalez17,franceschi22}). It can also affect the ionization rate and thus the magnetic activity (e.g., \citealt{lesur22pp7}). Finally, if the disc-to-star mass ratio is sufficiently high, (i.e. $q=M_{\rm disc}/M_{\star} \gtrsim 0.07 - 0.1$, \citealt{kratter16}), the disc self-gravity (hereafter SG) affects the morphology and the stability of the system. This may potentially lead to the development of gravitational instabilities (GI) in the form of a spiral structure \citep{jeans1902, linshu1964, Paczynski1978, binneytremaine1987, linpringle1987, bertinlin1996, kratter16}, which has been claimed to be observed (e.g., see the case of Elias 2-27, \citealt{perez16,huang18,paneque21}).
However, weighing protoplanetary discs is challenging. Indeed, the main disc gaseous component, molecular hydrogen (H$_2$), emits faintly in cold environments, thus making its direct detection extremely difficult. Consequently, one must rely on other gas mass tracers. A first approach consists in measuring fluxes of alternative gaseous tracers such as CO isotopologues (e.g., \citealt{miotello17}), $^{12}$CO being the most abundant one. Then, by assuming a model-dependent conversion factor from CO isotopologues to H$_2$, one estimates the total mass. Another viable tracer is dust: the gas mass can be estimated by measuring the dust mass from millimetre flux assuming the disc is optically thin and setting a value gas-to-dust ratio (e.g., 100 as in the ISM, \citealt{draine03}). However, all these methods rely on conversion factors that lack accurate constrains, and on the assumption that the observed dust is optically thin (for a more comprehensive review see \citealt{miotello22}). Another tracer that lately has been investigated is hydrogen deuteride (HD, e.g., \citealt{bergin13,trapman17}). When the disc vertical structure is known, a more reliable estimate of the disc mass can be obtained using HD, which tends to yield higher values compared to other tracers \citep{trapman17}. This method still lacks a large enough sample of measurements due to the absence of plans for a far-infrared observatory with the capability to observe HD. Recently, \cite{trapman22} investigated the possibility of using a combination of N$_2$H$^{+}$ and C$^{18}$O to determine the CO-to-H$_2$ ratio \citep{oberg23}, which however still depends on the assumed disc structure and N${_2}$ abundance. Lastly, a handful of additional approaches have been proposed in an attempt to provide a better characterization of this disc property. For example, this has been done by looking at multi-wavelength observations and comparing substructures to get an estimation of the Stokes number, and thus on the total disc mass or local surface density \citep{powell19,veronesi19}.
Also, by investigating the morphology of a kinematic feature linked to GI, known as GI-wiggle \citep{hall20,longarini21}, \cite{terry22} have recently found an approximately linear relationship between the amplitude of this wiggle and the disc-to-star mass ratio, thus obtaining an estimate for the mass in case of gravitationally unstable discs. 

Recently, an alternative approach to tracers or dust-to-H${_2}$ conversion has been proposed (\citealt{veronesi2021,lodato23}). It consists in measuring dynamically the mass of the disc. Indeed, when the disc is massive enough compared to the star, the gravity of the disc contributes considerably to the azimuthal gas velocity, making it super-Keplerian. In particular, the azimuthally averaged rotation curve is expected to transition from an inner Keplerian profile to an outer flatter one, indicative of the disc contribution. If high resolution kinematic data are available, the observed rotation curve can be fitted, thus obtaining a dynamical mass estimate. This method has been widely used in galaxies (e.g., \citealt{barbieri05}) and also applied to AGN discs (e.g., \citealt{lodato03}). It is also worth noting that it does not require precision on previous stellar mass measurements, since it simultaneously fits $M_\star$ and the disc mass, $M_{\rm d}$, from the rotation curve. Up to date, in the context of protoplanetary discs, the method has been applied to three systems, Elias 2-27 \citep{veronesi2021}, IM Lup and GM Aur \citep{lodato23}, providing for the first two a disc-to-star mass ratio ($\sim 0.1-0.4$) being consistent with an origin of the detected spirals as induced by GI.

The method, which is efficient for massive discs, relies on a simple model for the structure and the dynamics of the disc. As such, it is important to determine the range of disc-to-stellar mass up to which it can be applied with a sufficient accuracy. In this study, we provide a benchmark of the method by performing a set of Smoothed Particle Hydrodynamics (SPH) and radiative transfer simulations (\textsc{phantom}+\textsc{mcfost}, \citealt{price18, pinte09}), taking into account the gravity of the disc.

This article is organised as follows: in Section~\ref{sec:physicmodel} we describe the physical model use for the gas. In Section~\ref{sec:sims} we present and benchmark our set of numerical simulations. In Section~\ref{sec:method_result} we detail the analysis workflow and present our results. In Section~\ref{sec:uncertainties} we present our analysis on uncertainties. In Section~\ref{sec:discussion} we discuss them and highlight some possible limitations and future outlooks. We finally draw our conclusions in Section~\ref{sec:conclusion}.
\section{Physical model}\label{sec:physicmodel}

  We consider a smooth axisymmetric circular disc, and we neglect the effect of the magnetic field and dust feedback. Indeed, the first brings only small correction to the turbulence and magnetic pressure, while the latter may significantly modify the effect of the pressure gradient of the gas only for high dust-to-gas ratio, which is assumed not to be the case for our systems. We also do not consider discs with embedded planets. Low-mass planets would introduce a negligible correction of the order of $\propto M_{\rm p}/M_{\star}$. Massive planets would create gaps and modify the gas surface densities, but this is beyond the scope of this first exploration. Within these assumptions, the azimuthal velocity $v_\phi$ of a gas particle at a cylindrical distance $R$ and height $z$ from a central object of mass $M_\star$ is related to the gas pressure $P$ and the gravitational potential of the disc $\Phi_{\rm d}$ according to
\begin{equation}\label{eq:trc}
    v_\phi^2 = \frac{\mathcal{G} M_\star R^2}{(R^2+z^2)^{3/2}} + \frac{R}{\rho}\frac{\partial P(R,z)}{\partial R} + R\frac{\partial\Phi_{\rm d}}{\partial R}.
\end{equation}
The three contributions of the right hand-side of Eq.~\ref{eq:trc} correspond to the gravity of the central star, the radial pressure gradient and the self-gravity of the disc. We assume that the surface density of the gas follows a self-similar profile set by turbulent viscosity \citep{Lbell74}
\begin{equation}\label{eq:power-cutoff}
    \Sigma(R) = \frac{M_{\rm d}(2-\gamma)}{2\pi R_{\rm c}^2}\left(\frac{R}{R_{\rm c}}\right)^{-\gamma}\exp\left[-\left(\frac{R}{R_{\rm c}}\right)^{2-\gamma}\right],
\end{equation}
where $R_{\rm c}$ is the tapering radius, $M_{\rm d}$ the disc mass and $\gamma=1$ is the power-law index. Although this profile is commonly used for non self-gravitating discs, it can also be chosen in the self-gravitating case since the Keplerian shear is not affected by the disc SG. 
We also assume the disc to be vertically isothermal with the density profile given by
\begin{equation}\label{eq:volumedens}
    \rho(R,z)=\rho_0(R)\exp\left[-\frac{R^2}{H^2}\left(1-\frac{1}{\sqrt{1+z^2/R^2}}\right)\right],
\end{equation}
and the temperature profile to be a power law with the radius
\begin{equation}\label{eq:temperature}
    T(R) = T_{100}\left(\frac{R}{100\,\text{au}}\right)^{-q},
\end{equation}
where $H$ is the disc thickness, $\rho_0$ is the density at the midplane, and $T_{100}$ is a normalisation factor at 100 au. We stress that the discs considered are not only vertically, but also locally isothermal (see Sec.~\ref{sec:hydro}).

Following e.g. \cite{lodato23}, the rotation profile is
\begin{eqnarray}\label{model}
    v_{\phi}^2  = v_{\rm K}^2 \left\{1-\left[\gamma'+(2-\gamma)\left(\frac{R}{R_{\rm c}}\right)^{2-\gamma}\right]\left(\frac{H}{R}\right)^2 \right. \\ 
    \left. - q\left(1-\frac{1}{\sqrt{1+(z/R)^2}}\right) \right\} +v_d^2\,,\nonumber
\end{eqnarray}
where $\gamma^\prime = \gamma + 3/2 + q/2$, $ v_K^2 = {\mathcal{G}M_\star}/{R}$ and 
\begin{eqnarray}\label{sgterm}
\label{eq:sgpot}
v_{\rm d}^2 = \mathcal{G} \int_{0}^{\infty} \Bigg[ K(k)-\frac{1}{4}\left(\frac{k^{2}}{1-k^{2}}\right) \times  \\ 
\left(\frac{r}{R}-\frac{R}{r}+\frac{z^{2}}{R r}\right) E(k)\Bigg] \sqrt{\frac{r}{R}} k \Sigma \left(r\right) \mathrm{d} r\,. \nonumber
\end{eqnarray}
$K(k)$ and $E(k)$ are complete elliptic integrals and $k^2 = 4Rr/[(R + r)^2 + z^2]$ \citep{binneytremaine1987,bertin99}. Here the boundaries of the integral are chosen to corresponds to the inner and outer radius of the disc. The terms have been rearranged in four contribution: 1) a purely circular Keplerian term $v_{\rm K} \equiv \sqrt{\mathcal{G}M_{\star}/R}$, 2) a term related to the radial pressure gradient in the midplane, which gives a sub-Keplerian contribution, 3) another sub-Keplerian term originating from the pressure gradient and the stellar contribution in the vertical direction, and finally 4) the self-contribution of the disc (see Eq.~\ref{eq:sgpot}). To present some orders of magnitudes, the correction introduced by the vertical dependence of the stellar potential is $\sim (z/R)^2$ and is therefore of the same order as the one due to pressure gradient, and for this it should not be neglected (we derived with the updated model the disc mass of Elias 2-27 finding a result consistent with \citealt{veronesi2021}, see App.~\ref{app:elias}). In Appendix~\ref{app:resc}, these equations are provided in a dimensionless form for orders-of-magnitude estimates.

\begin{figure*}
    \centering
    \includegraphics[scale=0.5]{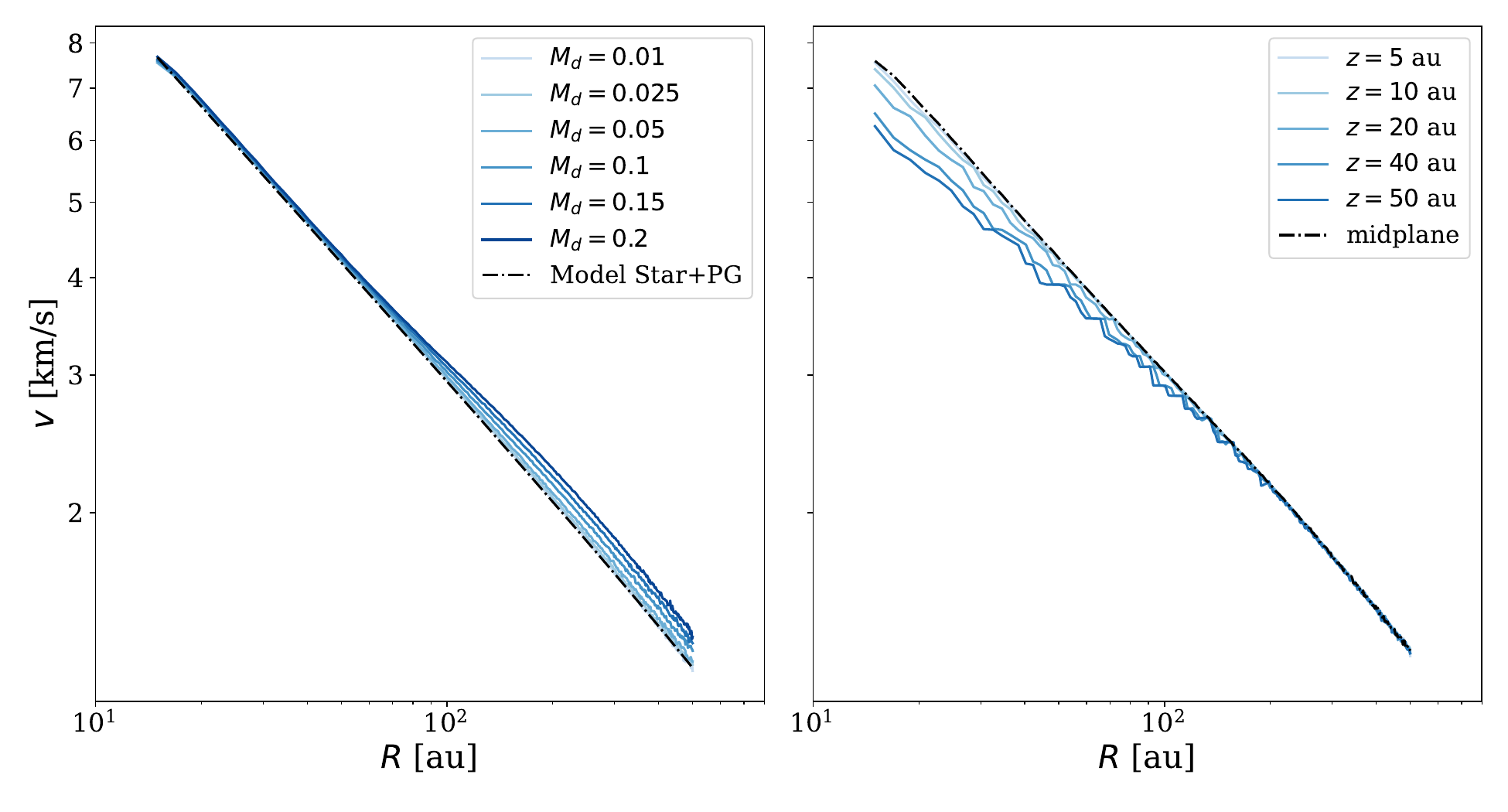}
  \caption{\textit{Left}: rotation curves in the midplane of the disc extracted from the SPH simulations with $H/R_{100\rm{au}}=0.075$ after 20 orbits at the outer disc radius, for different disc-to-star mass ratio (solid lines). The black dashed-dot line shows the rotation curve of the model given by Eq.~\ref{model}), including the pressure gradient term with zero disc mass. Except for profiles obtained with a disc-to-star mass ratio of 0.01, the rotation curves are distinct from a model without the disc self-gravity. \textit{Right}: same as left panel with $M_{\rm disc}=0.1\,M_{\odot}$, rotation curves being extracted for different heights (blue solid lines), and compared to the one obtained in the midplane ($z=0$, black dashed-dot line).}
    \label{hydrocurves_plot}
\end{figure*}

\section{Numerical simulations}\label{sec:sims}

\subsection{Hydrodynamical simulations}\label{sec:hydro}

We perform a suite of 3D Smoothed Particle Hydrodynamics (SPH) simulations of gaseous protoplanetary discs, using the code \textsc{phantom} \citep{price18}. The system consists of a central star of mass $M_\star=1\, M_{\odot}$ surrounded by a gas disc with mass $M_{\rm disc}=[0.01,0.025,0.05,0.1,0.15,0.2]\,M_{\odot}$. These simulations can be rescaled in terms of disc-to-star mass ratio (see the derivation in Appendix~\ref{app:resc}). The disc extends from $R_{\rm in}$ = 10 au to $R_{\rm out}$ = 300 au, and is simulated with $10^{6}$ SPH particles. The initial profile of surface density is a relaxed exponential tapered power law of Eq.~\ref{eq:power-cutoff}, with $R_{\rm c}=100$ au and $\gamma=1$. Simulations include the disc self-gravity as described in \citet{price18}, adopting a locally isothermal equation of state $P=c_{\rm s}^{2} \rho_{\mathrm{g}}$ with $q=0.5$ for the power-law index of the temperature radial profile (see Eq.~\ref{eq:temperature}). As a consequence of the absence of cooling 
, the morphology of the discs will be smooth and axisymmetric, without spiral density waves due to gravitational instability. This is consistent with the fact that to date, most of the observed discs are axisymmetric (with or without substructures, \citealt{ediskI}). Effects of self-gravity can be important even if the disc does not host spirals. The choice of the index for radial profile of temperature remains valid even if we would have taken into account dust in the hydrodynamical modeling, since variations in the temperature q-index are not expected to significantly impact the evolution of dust \citep{pinte14}.

The disc is vertically extended by initially setting up a disc aspect ratio of $(H/R)_{\rm c}=0.075$ with a Gaussian profile for the volume density (see Eq.~\ref{eq:volumedens}), ensuring nearly vertical hydrostatic equilibrium. These locally isothermal simulations do not require an extra cooling term. As such, we model the angular momentum transport throughout the disc using the SPH shock capturing viscosity \citep[see Sec. 2.6]{price18} with $\alpha_{\rm AV}\simeq0.19$, which results in a \citet{shakura73} viscous parameter $\alpha_{\mathrm{SS}} \approx 0.005$. The parameters used are given in App.~\ref{app:params_simu}.

\begin{table}
\caption{Masses, truncation radius and their uncertainties extracted from fitting hydrodynamical rotation curves with the model of Eq.~\ref{model}. Disc masses are all recovered with an accuracy of $\sim 5 - 12 \%$ for all disc-to-star mass ratios $[0.2,0.15,0.1,0.05,0.025,0.01]$, except for the extreme case $0.01$. The star mass $1\,M_{\odot}$ is always recovered ($~0.98-0.99\,M_{\odot}$). The input value for the truncation radius (100 au) is recovered with an accuracy of $\sim 6 - 20 \%$.}           \label{table:masses_hydro}     
\centering                          
 \resizebox{\columnwidth}{!}{
{\begin{tabular}{lcccc}        
\hline\hline       
 Sims & $M_{\rm d}$ [M$_\odot$] & $\Delta M_{\rm d}/M_{\rm d}$ & $R_{\rm c}$ [au] & $\Delta R_{\rm c}/R_{\rm c}$   \\  \hline\hline 
md0.2 &  0.19 & 0.05 & 119 & 0.19 \\  
md0.15 & 0.14 & 0.06 & 113 & 0.13 \\ 
md0.1 &  0.1 & 0. & 106 & 0.06\\ 
md0.05 &  0.05 & 0. & 94 & 0.06\\ 
md0.025 &  0.028 & 0.12 & 92 & 0.08 \\ 
md0.01 &  0.017 & 0.7 & 86 & 0.14\\ 
 \hline 
\end{tabular}           
}}
\end{table}

\subsection{Preliminary test: hydrodynamical rotation curves}

\begin{figure*}
    \centering
    \includegraphics[scale=0.3]{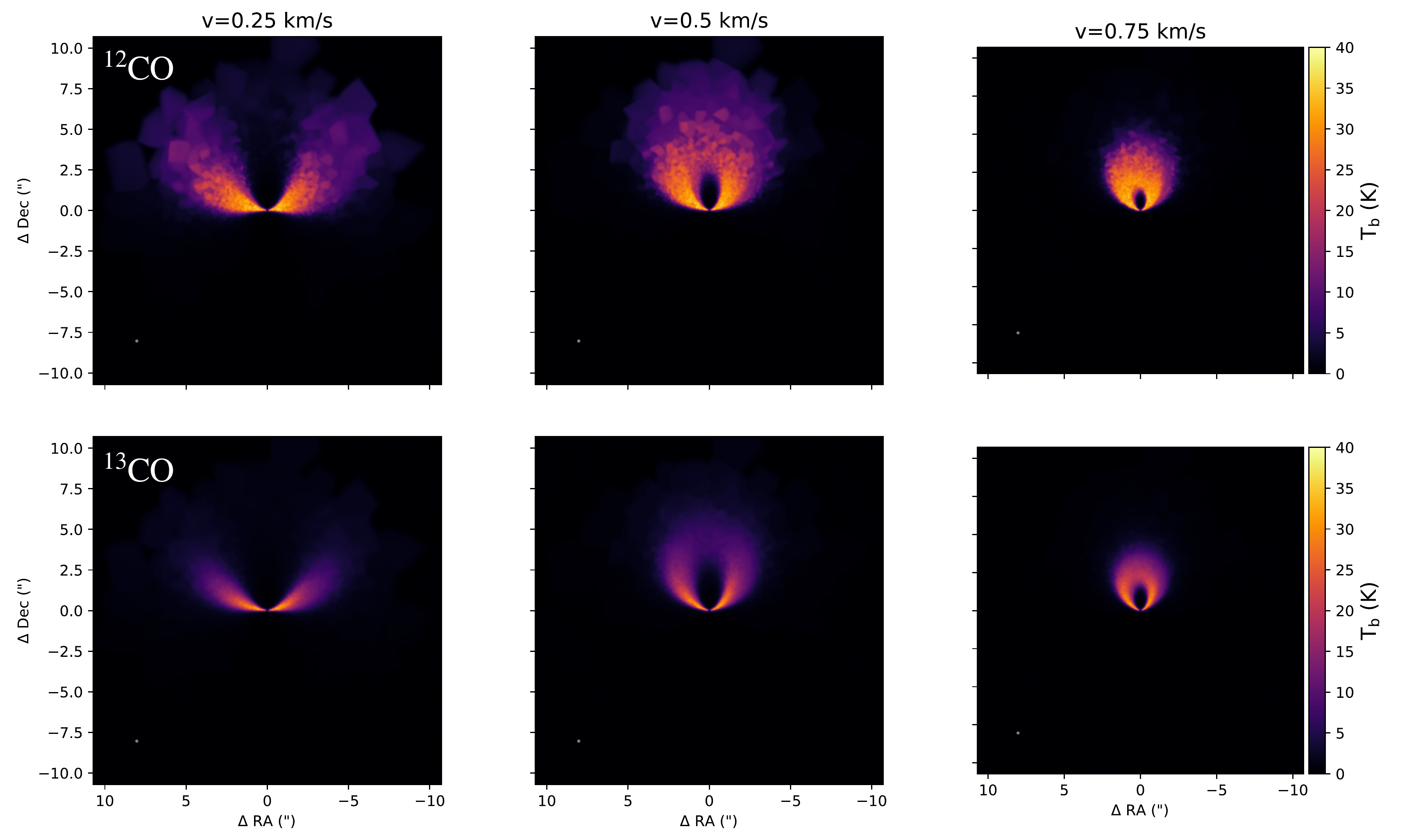}
    \caption{Example of channel maps obtained with \textsc{mcfost} for the $^{12}$CO (top row) and $^{13}$CO (bottom row) isotopologues, for a simulation with disc mass of $M_{\rm d}= 0.1\,M_{\odot}$. Different velocity channels are displayed, from 0.25 to 0.75 km/s (going from left to right). The chosen Gaussian beam used to convolve the image matches the value of MAPS survey (0.1''), and is showed with a grey circle in the left bottom corner of each channel maps. The velocity resolution is chosen to be 0.1 km/s (as in the MAPS survey).}
    \label{channelmaps}
\end{figure*}

As a first benchmark, we first extract the rotation curves directly from the hydrodynamical simulations
after having relaxed the disc to centrifugal equilibrium ($t\sim20\Omega_{\rm out}$, where $\Omega_{\rm out}$ is the orbital period of the disc outer radius). The rotation curves are obtained by interpolating azimuthal velocities of the SPH particles along the desired coordinates $(R,z)$, and then, azimuthally averaging the results. This step provides an uncertainty for the rotational velocity that is estimated as a standard deviation. Fig.~\ref{hydrocurves_plot} shows the rotation curves obtained by this procedure. The left panel shows the rotation profiles obtained in the midplane $(R, z=0)$, with disc masses ranging from 0.01 to 0.2 (blue lines). As a reference, the black line provides the rotation curve of the model given by Eq.~\ref{model}), including the pressure gradient term, but setting the mass of the disc to zero. As expected, the contribution of the disc self-gravity increases with the disc-to-star mass ratio. The right panel shows the rotation curves of the simulation with $M_\text{disc}=0.1$ at different height raging from $z=0$ to $z=50$ au. As expected from the model, the difference of azimuthal velocity at different heights decreases with radius.

To benchmark the validity of the formulation given by Eq.~\ref{model}, we fit the extracted rotation curves with the code \textsc{dysc}\footnote{\url{https://github.com/crislong/DySc}}, that implements the model described in Sec.~\ref{sec:physicmodel}. Results are shown in Table~\ref{table:masses_hydro}. The fitting procedure relies on a Markov Chain Monte Carlo using the python package \textsc{emcee} \citep{emcee}. We use a Gaussian likelihood with flat priors. The code is fitting simultaneously both the masses of the star and the disc, as well as the truncation radius $R_c$, similarly to the procedure described in \citet{lodato23}. After this step, we recover the mass of the disc with an uncertainty of $\sim 5-12\%$ (see Table~\ref{table:masses_hydro}) for disc-to-star mass ratios ranging between $0.025$ and $0.2$. in those cases, discs are sufficiently massive for the self-gravity to have measurable contribution. For a disc-to-star mass ratio of $0.01$, the disc mass cannot be extracted accurately even from the numerical simulations. 

\begin{figure}
    \centering
    \includegraphics[width=0.5\textwidth]{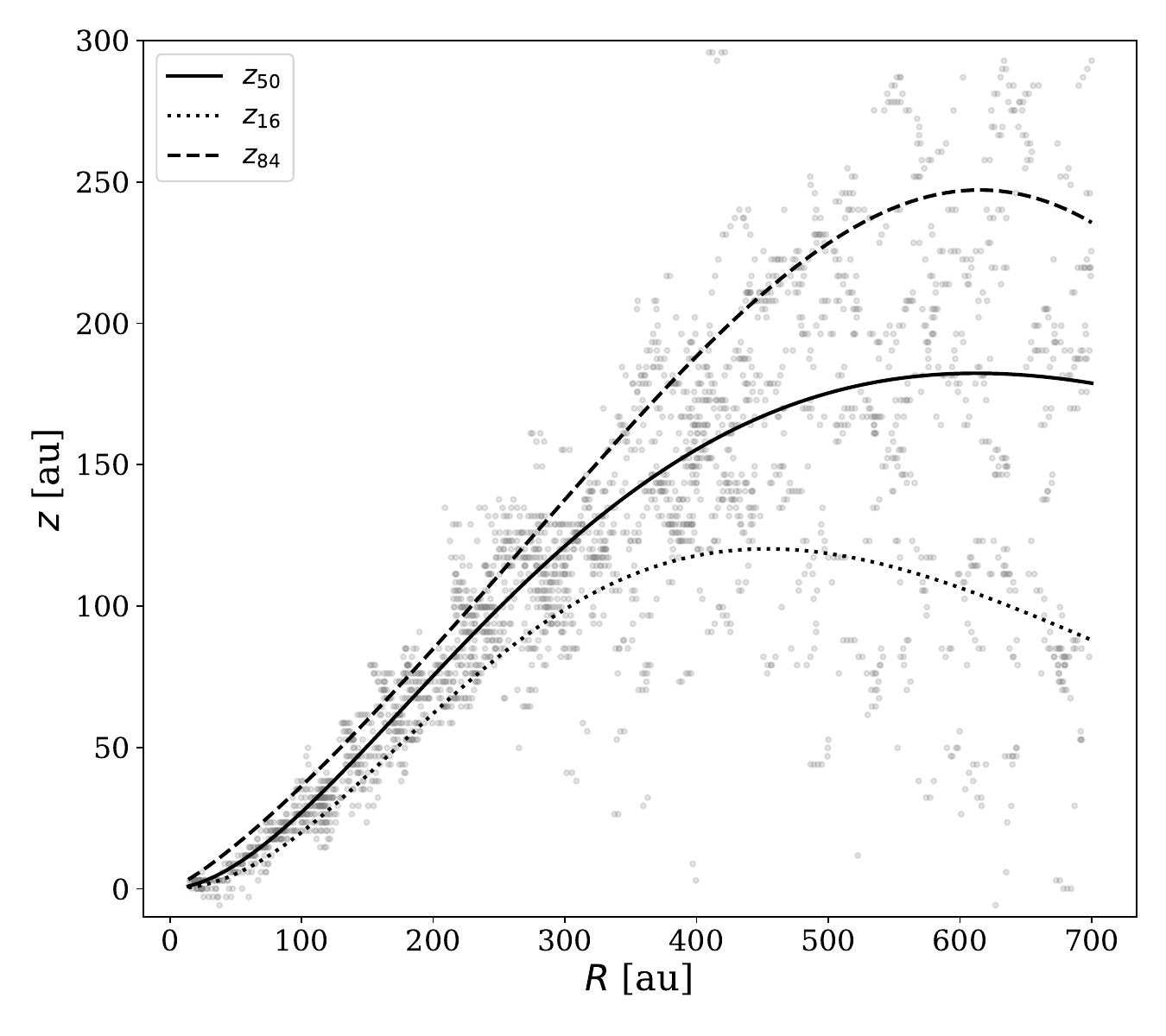}
    \caption{Example ($^{12}$CO, $M_{\rm disc}=0.1\,M_{\odot}$) of one of the emitting layer derived with \textsc{disksurf} (grey dots) and then fitted with an exponentially tapered power-law (solid line). We also show the 16th (dotted line) and the 84th (dashed line) percentiles, which we take into account to compute the rotation curve errors.}
    \label{disksurf_em_surface}
\end{figure}
\subsection{Radiative transfer and synthetic observations}\label{sec:radtrans}

We then compute the thermal structure and synthetic observations of our numerical discs by performing 3D radiative transfer simulations with the code \textsc{mcfost} \citep{pinte06,pinte09}. We use a Voronoi tesselation where each \textsc{mcfost} cell corresponds to a gas SPH particle. The goal is to generate mock $^{12}$CO and $^{13}$CO isotopologue channel maps, from which we will extract disc rotation curves. 

The main inputs for the radiative transfer modelling are the source of luminosity, the gas density profile extracted from the simulations, and models for dust opacities and densities. Self-gravitating isothermal discs do not present any kind of sub-structures and the underlying dust density profile can therefore assumed to be smooth. We do not account for an eventual dust drift, since Stokes numbers are expected to be small in these discs, and we consider short-enough disc evolutions. The dust contribution to the gravitational potential of the disc is also negligible. As such, we adopt a constant dust-to-gas ratio of$\sim 10^{-2}$ that corresponds to a standard averaged ISM \citep{draine03}.

The thermal structure of the disc is computed with the following assumptions. At first, emission is at local thermal equilibrium and $T_{\rm gas} = T_{\rm dust}$. 
Dust is treated as a mixture of silicates (70\%) and carbon (30\%) \citep{draine84}, and the optical properties are calculated using Mie theory for spheres \citep{andrews09}. Opacities are computed following the procedure described in the DIANA model \citep{woitke16,min16}. We assume an ISM-like grain size distribution ($dn/da \propto a^{-3.5}$), with $a_{\rm min}=0.01\,\mu$m and $a_{\rm max}=1$ mm. The disc is heated passively, i.e. the source of radiation is only the central star, which is assumed to radiate isotropically with a Kurucz spectrum at 4000 K. The expected channel maps are computed via ray-tracing, using $10^7$ photon packets to sample the radiation field.
The disc inclination with respect to the line of sight is $i = 30^{\circ}$, and the system is simulated to be located at 140 pc, which corresponds to a typical protostellar discs in a star-forming region such as Taurus. For $^{12}$CO and $^{13}$CO, we consider the J=2-1 transition and assume abundances of $10^{-4}$ and $1.4 \times 10^{-6}$ respectively. \textsc{mcfost} simulations are post-processed with \textsc{pymcfost} \footnote{\url{https://github.com/cpinte/pymcfost}}, by simulating a velocity resolution of $0.1\,\rm{km/s}$. We then spatially convolve the channels with a Gaussian beam of 0.1 arcsec, similarly to the value of the MAPS survey \citep{maps21}. We finally introduce Gaussian noise with an RMS of $5\text{mJy/beam}$.

Fig.~\ref{channelmaps} shows an example of the channel maps obtained for a simulation with a disc mass of $0.1\,M_{\odot}$. The $^{12}$CO channels appears more radially extended compared to $^{13}$CO channels. 
$^{13}$CO channels are also less patchy, since the $\tau = 1$ region is located closer to the midplane, a region with higher resolution. As a result the $^{13}$CO disc rotation curves will be less noisy than the curves extracted from the $^{12}$CO channels (see Fig.~\ref{rotationcurves_plot}). 
We note that while this is the case for simulations, in real observations we do not expect $^{13}$CO rotation curves to be less noisy than the $^{12}$CO. Indeed, $^{13}$CO is much less abundant than $^{12}$CO, and at the same sensitivity, the S/N in $^{13}$CO should be lower than for $^{12}$CO.

\section{Method and results}\label{sec:method_result}

\subsection{Workflow description} \label{sec:workflow}
From the cubes of synthetic data, we first determine the height of the emitting layer with \textsc{disksurf} \citep{disksurf}. The emitting layer is defined as the region where the emission we observe originates. Isotopologues have specific optical depth and column density, and as such, the emitting layer can be located farther or closer to the midplane of the disc. Having a precise estimate of the height of the emitting layer is crucial to correctly deproject azimuthal velocities and compute the model Eq.~\ref{model} at the correct height.

We then use \textsc{eddy} \citep{eddy} to extract the rotation curves. Two methods have currently been developed to retrieve the observed velocity (specifically to fit the line centroids): the quadratic and the Gaussian methods. The quadratic fit aims at fitting a quadratic curve to the brightest pixel on either side of the maximum of the pixel.
It depends on the velocity sampling, and it is sensitive to the channel correlation. On the other hand, for the Gaussian method, which finds the line center by fitting a Gaussian profile to it, we note that the selected velocity range may affect the result in case of skewed profiles. 

To perform our analysis we chose the quadratic method, and we will briefly discuss this choice in Sec.~\ref{sec:extrac_v} (for an extensive discussion of the methods, see e.g. \citealt{lodato23} and \citealt{Teague_2018_methods}). 

Finally, we fit the rotation curves with the model of Eq.~\ref{model} using the code \textsc{dysc} (see Sec.~\ref{sec:physicmodel}).

\subsection{Retrieving the height of the emitting layer}\label{sec:z(r)}

\begin{table}\caption{$^{12}$CO and $^{13}$CO fit results for the disc emitting layer relevant parameters ($z_0$, $\psi$, $R_{\rm t}$, $q_{\rm t}$) obtained with \textsc{disksurf} from the 50th percentiles of the particle vertical distribution.}
\label{tab:emitting}
\resizebox{\columnwidth}{!}
{
\begin{tabular}{lcccccc} 
  & md0.01 & md0.025 & md0.05 & md0.1 & md0.15 & md0.2\\ \hline \hline
 $^{12}$CO\Tstrut\Bstrut  &  &  & &  & &  \\  
 \hline
  \begin{tabular}[l]{@{}l@{}} $z_0$\\ $\psi$\\ $R_{\rm t}$\\ $q_{\rm t}$\end{tabular} & \begin{tabular}[l]{@{}l@{}} 0.83 \\ 3.06 \\ 0.49 \\ 0.69 \end{tabular} \vspace{1mm}& \begin{tabular}[l]{@{}l@{}} 0.24 \\ 1.88 \\ 3.7 \\ 1.77 \end{tabular} & \begin{tabular}[l]{@{}l@{}} 0.38 \\ 1.66 \\ 3.54 \\ 1.46 \end{tabular}  & \begin{tabular}[l]{@{}l@{}} 0.45 \\ 1.85 \\ 2.76 \\ 1.11 \end{tabular}  & \begin{tabular}[l]{@{}l@{}} 0.33 \\  1.13 \\  5.1 \\ 9.47 \end{tabular}  & \begin{tabular}[l]{@{}l@{}} 0.25 \\ 1.87 \\ 4.31 \\ 2.35 \end{tabular} \\ 
\hline 
 $^{13}$CO\Tstrut\Bstrut  &  &  & &  & &  \\
 \hline 
  \begin{tabular}[l]{@{}l@{}} $z_0$\\ $\psi$\\ $R_{\rm t}$\\ $q_{\rm t}$\end{tabular} & \begin{tabular}[l]{@{}l@{}}  0.078 \\ 1.25 \\ 3.01  \\ 6.27 \end{tabular} & \begin{tabular}[l]{@{}l@{}} 0.09 \\ 1.52 \\ 3.25 \\  5.39 \end{tabular} & \begin{tabular}[l]{@{}l@{}} 0.097 \\ 1.72 \\ 3.35 \\ 4.22 \end{tabular} & \begin{tabular}[l]{@{}l@{}} 0.09 \\ 1.82 \\ 3.46 \\ 3.53 \end{tabular}  & \begin{tabular}[l]{@{}l@{}} 0.09 \\ 2.04 \\ 3.38 \\ 2.83 \end{tabular}  & \begin{tabular}[l]{@{}l@{}} 0.09 \\ 2.02 \\ 3.33 \\ 2.61 \end{tabular} \\ 
\hline 
\end{tabular}           
}
\end{table}

We first use the code \textsc{disksurf} \citep{disksurf} to extract the height of the emitting layer $z^{n}\left(r_{i} \right)$ of the $i$-th radial bin for the $n$-th channel. \textsc{disksurf} implements the method presented in \cite{Pinte18}. Briefly, this is a geometrical method that allow to trace the emission maxima of each velocity channel and reconstruct the position of the CO layers. For each bin, we determine the heights $z^{n}_{16}\left(r_{i} \right)$,  $z^{n}_{50}\left(r_{i} \right)$ and  $z^{n}_{84}\left(r_{i} \right)$ that corresponds to the 16-th, 50-th and 84-th percentile of the distribution respectively. We then use the tapered power-law 
\begin{equation}\label{z_r_eq}
    z(R) = z_0\left[\text{arcsec}\right] \left(\frac{R}{1\text{arcsec}}\right)^\psi \exp\left[-\left(\frac{R}{R_t}\right)^{q_t}\right],
\end{equation}
to parametrize continuously the emitting surfaces $z_{16}\left(r \right)$, $z_{\rm em}\left(r\right) = z_{50}\left(r \right)$ and $z_{84}\left(r \right)$. The parameters obtained are used to retrieve the height of the molecular emission surface $z_{\rm em}\left(r\right)$ as well as to estimate the uncertainties associated to the procedure. 
\begin{figure*}
    \centering
    \includegraphics[scale = 0.47]{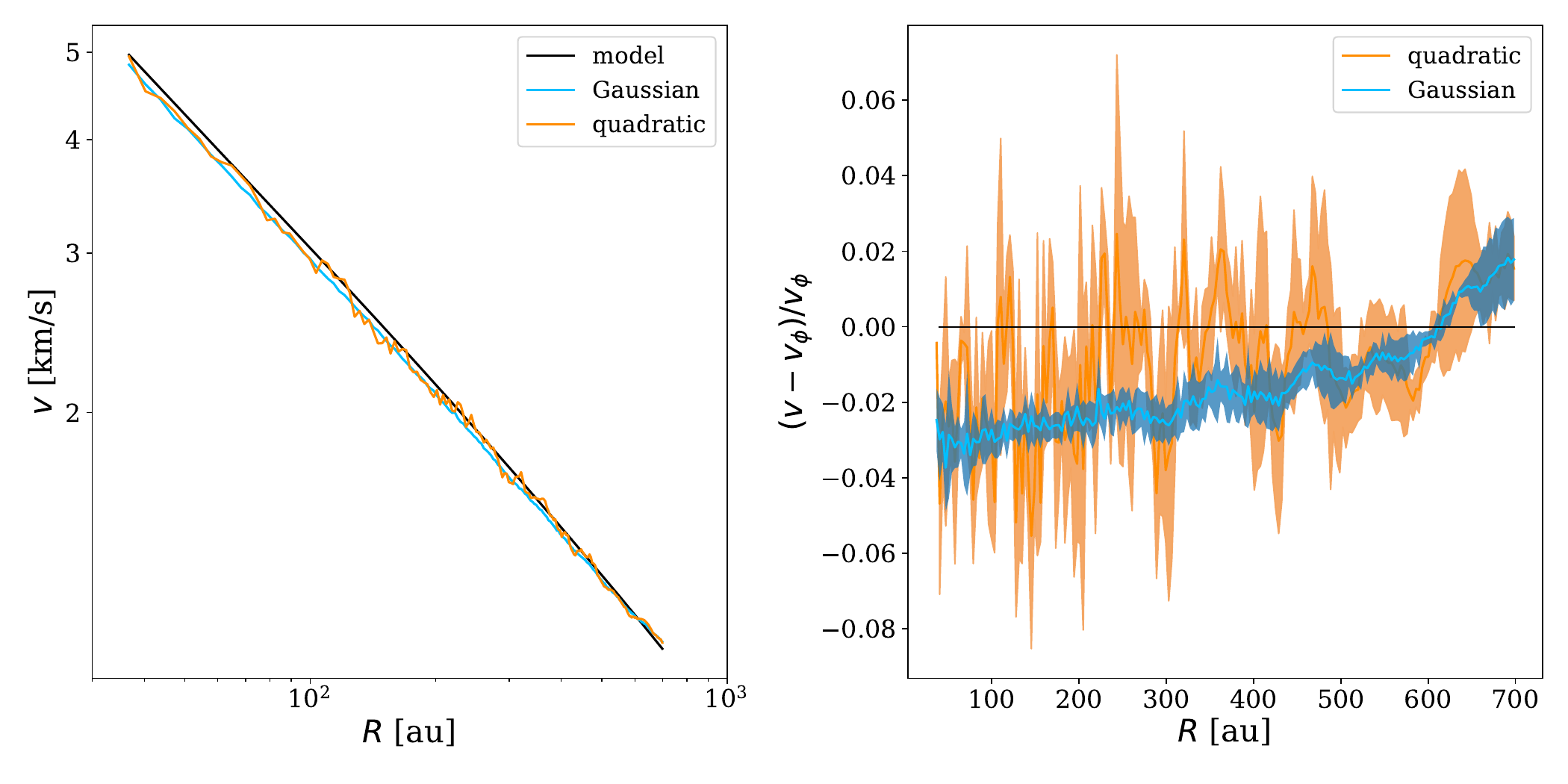}
    \caption{Differences between rotation curves obtained with the Gaussian and quadratic methods. Left panel: comparison between the rotation curves from the model (black line, Eq.~\ref{model} with a disc mass $M_{\rm d}=0.1\,M_{\odot}$), the Gaussian method (blue line) and the quadratic method (orange line). We observe that the Gaussian curve is systematically shifted with respect to the model. Right panel: difference between the extracted rotation curve (Gaussian method in blue, quadratic method in orange) and the model $v_{\phi}$ from Eq.~\ref{model}). Gaussian extraction is biased by the lower surface, shifting the curve towards lower velocity in most of the radial extent of the disc. Conversely, the quadratic method, although noisier, better reproduces the theoretical model.}
    \label{quadratic_gaussian}
\end{figure*}

Table~\ref{tab:emitting} presents the results obtained for $z_{\rm em}\left(r\right)$ that will be used as reference surface through the following analysis. As an example, Fig.~\ref{disksurf_em_surface} shows the emission surface (grey markers) obtained with \textsc{disksurf} for the $^{12}$CO datacube, from a mock observations generated by a simulation with $M_{\rm disc}=0.1\,M_{\odot}$. The three lines are the fits of the data points, namely $z_{\rm em} = z_{50}$ (solid line), as well as $z_{16}$ (dotted line) and $z_{84}$ (dashed line). The scatter between the three fitted surfaces is important with this procedure : it is hence crucial to keep the information of this dispersion when analysing uncertainties on the rotation curve. 

As expected, we also find that there is a clear trend between the height of the emitting layer and the disc mass: more massive discs have higher emitting layer, since their surface density is higher and they are optically thicker. Also, the height of the $^{13}$CO is lower with respect to the $^{12}$CO because of its reduced abundance, tracing a region closer to the midplane. This is shown more in detail in Figs.~\ref{surface_mass} and ~\ref{surface_13co_12co}). 

Finally, for consistency, we decide not to includeat this stage in our analysis the simulation with $M_{\rm d}/M_\star=0.2$. Indeed, in this case the $^{12}$CO emission is poorly reproduced since the SPH resolution at the emitting surface is not sufficient (see Sect.~\ref{sec:res} for a resolution analysis).

\begin{figure*}
    \centering
    \includegraphics[scale = 0.5]{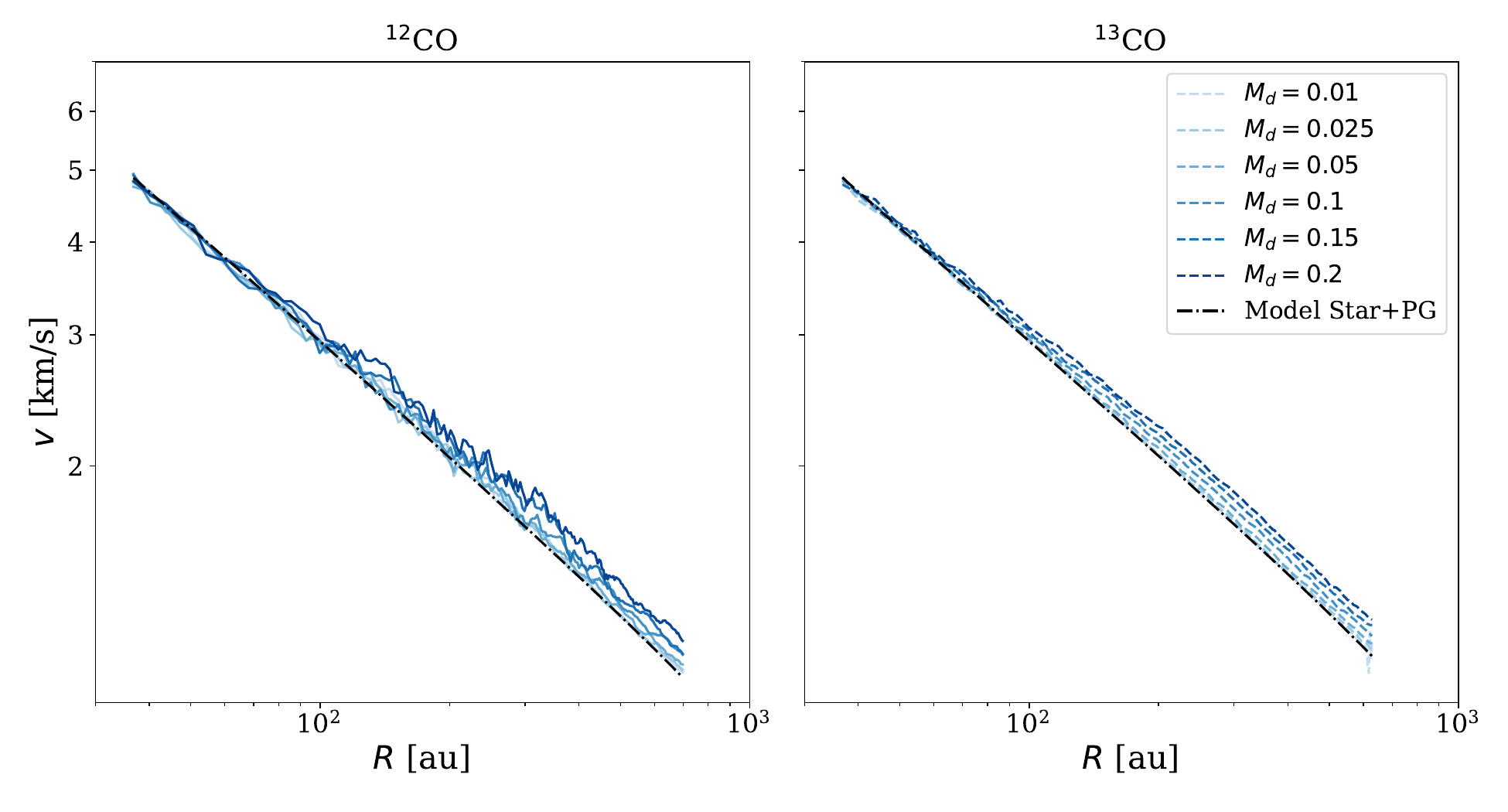}
    \caption{Quadratic rotation curves obtained for all the disc masses simulation (from $M_{\rm disc}=0.01\,M_{\odot}$ to $M_{\rm disc}=0.2\,M_{\odot}$) for the $^{12}$CO isotopologue (solid lines in the left panel) and the $^{13}$CO isotopologue (dashed lines in the right panel). The black dashed-dot line shows the analytical rotation curve obtained by only considering the star and pressure gradient contribution. } 
    \label{rotationcurves_plot}
\end{figure*}

\subsection{Extracting the rotation curve}\label{sec:extrac_v}

We extract the azimuthal velocity with \textsc{eddy} \citep{eddy} using the harmonic oscillator method, following \cite{teague18} and \cite{teague18b}. Further details about the extraction methods can be found in Section 3.3.1 of \cite{lodato23}. 

For each emitting layers $z_{16},z_{\rm em} = z_{50}$ and $z_{84}$, we compute three rotation curves $v_{16}$, $v_{50}$ and $v_{84}$. We then assume the azimuthal velocity of the system to be $v_{\rm em} = v_{50}$, with an uncertainty $\sigma_v$ estimated to be 
\begin{equation}
    \sigma_v = \sqrt{|v_{84} - v_{16}|^2 + \sigma_\text{eddy}^2}\,,
\end{equation}
where $\sigma_\text{eddy}$ is the numerical error obtained with \textsc{eddy}. This procedure refines the approach of \citet{lodato23} by including uncertainties associated to estimate of the height the emitting layer. 

Since we fixed the inclination of the disc to be $i=30^\circ$, we use the Quadratic method to extract the rotation curve, that fits only the peak of emission, which is less sensitive to the lower surface. Indeed, using instead the Gaussian method provides the following bias: the emission coming from the lower surface systematically shifts the position of the line centroids, resulting in a systematic error for the velocity estimate. Fig.~\ref{quadratic_gaussian} shows a comparison between the quadratic and the Gaussian methods. Interestingly, despite being smoother, the curve obtained with the Gaussian method underestimates the velocity in the inner disc and overestimate it in the outer disc. This happens because the method tries to fit a double-peaked spectrum with a single Gaussian. 
We note that in this work we test these two methods, and despite both of them being biased, we choose the quadratic method since the Gaussian one presents an overall larger systematic shift of the velocity. Moreover, already with the quadratic method we manage to recover the correct disc and star masses. For completeness, we also tested our workflow using a rotation curve retrieved with a double Gaussian fit. However, the double-Gaussian method fails retrieving the correct velocity in the outer region of the disc ($R>300$ au) where the S/N is too low (it fits the noise and not the signal, see App.~\ref{app:doublegauss}).

Fig.~\ref{rotationcurves_plot} shows the rotation curves obtained with the quadratic method following the procedure above, for both CO isotopologues ($^{12}$CO in the left panel, $^{13}$CO in the right panel). We compare them with the analytical rotation curve obtained assuming a disc of zero mass (black dashed-dot line). As expected, the rotation curve derived with this approximation fits only the disc with the lower mass ($M_{\rm d}=0.01\,M_{\odot}$), since this is the only case where the disc contribution can be considered as negligible. This will be confirmed in Sec.~\ref{sec:results_fit}.

\subsection{Fitting procedure}\label{sec:fit}
For every simulation, we obtain two rotation curves: one for $^{12}$CO and another for $^{13}$CO. These curves are then fitted using 
the self-gravitating model of Eq.~\ref{model}. The free parameters are the star and disc mass, and the disc truncation radius. The fits are performed with an MCMC algorithm as implemented in \textsc{emcee} \citep{emcee}. We choose a simple Gaussian likelihood, and flat priors respectively $[0,5]\text{M}_\odot$ for the star mass\footnote{We also tried using a prior with a narrower range around $1\,M_{\odot}$ and the result does not change.}, $[50,300$]au for the truncation radius and $[0, 0.5]\text{M}_\odot$ for the disc mass. We choose 250 walkers and 1000 steps (having verified that convergence is reached). We fit the two isotopologues both individually, and then simultaneously. 

\begin{table*}\caption{Results for the $^{12}$CO, $^{13}$CO and the combined fit procedure. True values are: $M_{\star}=1\,M_{\odot}$, $M_{\rm d}=[0.01,0.025,0.05,0.1,0.15,0.2]\,M_{\odot}$, $R_{\rm c}=100$ au. Uncertainties ($\Delta X/X$) are also shown and they are discussed in Section~\ref{sec:accuracy}.} \label{tab:fit}
\centering
{\begin{tabular}{lllllll}
 & $^{12}$CO &  $\Delta X/X_{^{12}{\rm CO}}$ & $^{13}$CO &  $\Delta X/X_{^{13}{\rm CO}}$& Combined  & $\Delta X/X_{\rm comb}$\\ \hline\hline
md0.01 & \begin{tabular}[c]{@{}l@{}}$M_\star=1.02$\\ $M_d=0.03$\\ $R_c=80.00$\end{tabular} &  \begin{tabular}[c]{@{}l@{}} 0.02 \\ 1.99 \\ 0.2 \end{tabular} & \begin{tabular}[c]{@{}l@{}}$M_\star=0.99$\\ $M_d=0.00$\\ $R_c=105.15$\end{tabular} &  \begin{tabular}[c]{@{}l@{}} 0.01 \\ 1.0 \\ 0.05 \end{tabular}  & \begin{tabular}[c]{@{}l@{}}$M_\star=0.99$\\ $M_d=0.00$\\ $R_c=103.7$\end{tabular}  & \begin{tabular}[c]{@{}l@{}} 0.01 \\ 1.0 \\ 0.037 \end{tabular} \\ 
\hline md0.025 & \begin{tabular}[c]{@{}l@{}}$M_\star=0.99$\\ $M_d=0.04$\\ $R_c=92.17$\end{tabular} &  \begin{tabular}[c]{@{}l@{}} 0.01 \\ 0.6 \\ 0.078 \end{tabular}  & \begin{tabular}[c]{@{}l@{}}$M_\star=0.99$\\ $M_d=0.00$\\ $R_c=115.55$\end{tabular} &  \begin{tabular}[c]{@{}l@{}}  0.01 \\ 1.0 \\ 0.15 \end{tabular}  & \begin{tabular}[c]{@{}l@{}}$M_\star=0.99$\\ $M_d=0.00$\\ $R_c=115.2$\end{tabular}  & \begin{tabular}[c]{@{}l@{}} 0.01 \\ 1.0 \\ 0.15 \end{tabular} \\ 
\hline md0.05 & \begin{tabular}[c]{@{}l@{}}$M_\star=0.99$\\ $M_d=0.044$\\ $R_c=102.78$\end{tabular} &  \begin{tabular}[c]{@{}l@{}} 0.01 \\ 0.12 \\ 0.028 \end{tabular} & \begin{tabular}[c]{@{}l@{}}$M_\star=0.97$\\ $M_d=0.07$\\ $R_c=94.27$\end{tabular} &  \begin{tabular}[c]{@{}l@{}}  0.03 \\ 0.4 \\ 0.057 \end{tabular} & \begin{tabular}[c]{@{}l@{}}$M_\star=0.98$\\ $M_d=0.055$\\ $R_c=97.8$\end{tabular}  & \begin{tabular}[c]{@{}l@{}} 0.02 \\ 0.099 \\ 0.022 \end{tabular} \\
\hline md0.1 & \begin{tabular}[c]{@{}l@{}}$M_\star=1.04$\\ $M_d=0.09$\\ $R_c=88.33$\end{tabular} &  \begin{tabular}[c]{@{}l@{}} 0.04 \\ 0.10 \\ 0.117 \end{tabular} & \begin{tabular}[c]{@{}l@{}}$M_\star=0.97$\\ $M_d=0.12$\\ $R_c=90.8$\end{tabular} &  \begin{tabular}[c]{@{}l@{}} 0.03 \\ 0.20 \\ 0.09\end{tabular}  & \begin{tabular}[c]{@{}l@{}}$M_\star=0.97$\\ $M_d=0.12$\\ $R_c=91.2$\end{tabular}  & \begin{tabular}[c]{@{}l@{}} 0.03 \\ 0.19 \\ 0.088 \end{tabular} \\ 
\hline md0.15 & \begin{tabular}[c]{@{}l@{}}$M_\star=1.00$\\ $M_d=0.18$\\ $R_c=86.00$\end{tabular} &  \begin{tabular}[c]{@{}l@{}} 0.0 \\ 0.2 \\ 0.14 \end{tabular} & \begin{tabular}[c]{@{}l@{}}$M_\star=1.00$\\ $M_d=0.15$\\ $R_c=87.5$\end{tabular} &  \begin{tabular}[c]{@{}l@{}} 0.0 \\ 0.0 \\ 0.125 \end{tabular}  & \begin{tabular}[c]{@{}l@{}}$M_\star=1.00$\\ $M_d=0.15$\\ $R_c=88.114$\end{tabular}  & \begin{tabular}[c]{@{}l@{}} 0.0 \\ 0.0 \\ 0.12 \end{tabular} \\ 
\hline md0.2 & \begin{tabular}[c]{@{}l@{}}$M_\star=1.1$\\ $M_d=0.165$\\ $R_c=87.26$\end{tabular} &  \begin{tabular}[c]{@{}l@{}} 0.1 \\ 0.175 \\ 0.13 \end{tabular}  & \begin{tabular}[c]{@{}l@{}}$M_\star=1.06$\\ $M_d=0.15$\\ $R_c=84.9$\end{tabular} &  \begin{tabular}[c]{@{}l@{}} 0.06 \\ 0.25 \\ 0.15\end{tabular} & \begin{tabular}[c]{@{}l@{}}$M_\star=1.12$\\ $M_d=0.09$\\ $R_c=96.23$\end{tabular}  & \begin{tabular}[c]{@{}l@{}} 0.12 \\ 0.55 \\ 0.037  \end{tabular} \\ \hline
\end{tabular}
}
\end{table*}

\subsection{Fit results}\label{sec:results_fit}
Table~\ref{tab:fit} collects the values retrieved for the star mass $M_{\star}$, the disc mass $M_{\rm d}$ and the disc truncation radius $R_{\rm c}$, for simulations with disc masses in ranging  within $M_{\rm d} = 0.01-0.2\,M_{\odot}$, and a disc aspect ratio of 0.075. The disc-to-star mass ratio threshold below which the expected value can not be recovered is $0.05$. For resolution issues, we also exclude the simulation with $M_{\rm d} = 0.2\,M_{\odot}$ (see Sec~\ref{sec:z(r)}). From our procedure, we are able to measure a non-zero value for the disc mass ($M_{\rm disc} \geq 0.05\,M_{\odot}$), which indicates that the model for the rotation curve should include the term from the gravity of the disc.

\section{Uncertainties}\label{sec:uncertainties}
\begin{figure*}
    \centering
    \includegraphics[scale=0.32]{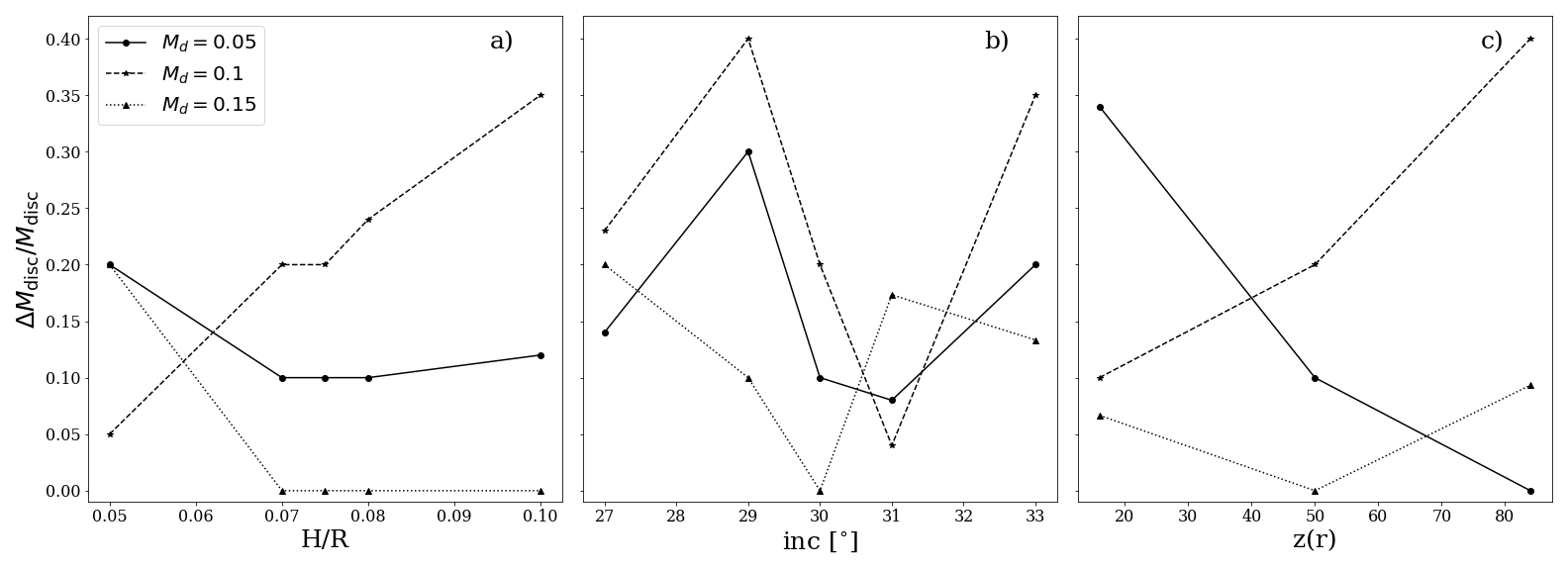}
    \caption{Uncertainties related to the fitting procedure for the disc mass $M_{\rm d}$. Estimates are given as functions of the aspect ratio of the disc $H/R$ (left column), the disc inclination (middle column) and the disc emitting layer $z(R)$ (right column). Different markers and lines styles represent results obtained for simulations with different disc masses.}
    \label{fig:accuracy_mass}
\end{figure*}

\subsection{Physical parameters}\label{sec:accuracy}

In the treatment of real data, the three principal sources of uncertainties are the height of the emitting layer, the aspect ratio the disc and its inclination. Uncertainties on $z\left( r \right)$ are estimated with the values of $z_{16}$ and $z_{84}$ previously found. To estimate uncertainties associated to $H/r$ and $i$, we first note that in the fitting procedure of Sect.~\ref{sec:method_result}, we enforce the values of $H/r$ and $i$ that were set up in the numerical simulation. We therefore estimate uncertainties associated to those parameters by carrying over the fit over the same synthetic channel maps while varying $H/r$ and $i$ over a range of values. For $H/r$, we simply perform new fits over the rotation curves previously extracted. For a set up value of $H/R = 0.075$, we test $H/R$ = [0.05, 0.07, 0.08, 0.1]. For inclinations, uncertainties estimates require the extraction of new disc emitting layers, implying to obtain novel rotation curves. For a set up value of $i = 30^{\circ}$, we test $i$ = [27, 29, 31, 33]$^{\circ}$. The values of the uncertainties obtained by this procedure are summarised in Table.~\ref{tab:fit}. Fig.~\ref{fig:accuracy_mass} shows these uncertainties for discs whose mass can be accurately recovered by the fitting procedure. This excludes the md0.01 and md0.025 (small masses) and md0.2 (lack of resolution). A key result of this study is that disc masses of self-gravitating discs can be estimated from channel maps with averaged systematic uncertainties of order $\sim 25\%$. The three parameters $H/r$, $i$ and $z$ have similar contributions as sources of errors. No clear trend appears when varying $H/r$ and $z$. Values that differ from the expected one still provide a mass estimates with a same level of uncertainty. On the other hand precise values of $i$ provide uncertainties of order $5 - 10\%$ whereas an error of a few degrees provide uncertainties of order $20 - 30\%$. Hence, the ability to accurately determine disc masses by our procedure is quite reliable. For instance, when recovering a disc mass of $0.1\,M_{\odot}$, the estimated range spans from 0.075 to 0.125. As a remark, one obtains simultaneously $\Delta M_{\star}/M_{\star} \sim 8$\% for an inclination error of $\pm 3^{\circ}$;
and $\Delta R_{\rm c}/R_{\rm c} \sim 15$\% for a reasonable disc aspect ratio range of $\pm 0.025$ (see App.~\ref{app:others} for a discussion).

\begin{figure}
    \centering
    \includegraphics[scale=0.43]{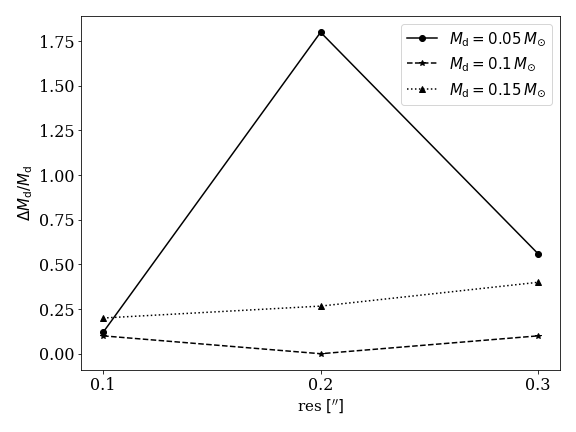}
    \caption{Uncertainties obtained from our fitting procedure for the disc mass $\Delta M_{\rm d}/M_{\rm d}$ for the combined fit procedure with different spatial resolutions [0.1,0.2,0.3]''. Different markers (and lines) represent results obtained for simulations with different disc masses.}
    \label{fig:accuracy_res}
\end{figure}

\subsection{Spatial resolutions}
In order to test how the minimum detectable mass is affected by the spatial resolution we produce new synthetic images by convolving the channel maps previously obtained in Sec.\ref{sec:radtrans} with different Gaussian beams sizes [0.2, 0.3]''. 
We then extract new rotation curves with the procedure described in Sec.~\ref{sec:z(r)} and~\ref{sec:extrac_v}, and we fit them with the \textsc{dysc} code (Sec.~\ref{sec:fit}). In Fig.~\ref{fig:accuracy_res} we report results obtained for the disc mass (different markers and lines), showing uncertainties obtained as a function of the spatial resolution of the observations. When $M_{\rm d}= 0.05\,M_{\odot}$, the uncertainty ranges from $\sim50-100\%$, while for the other two cases, $M_{\rm d}= [0.1-0.15]\,M_{\odot}$, it ranges within $\sim30-40\%$, which is larger compared to the 0.1'' case previously discussed ($\sim25\%$, see Sec.~\ref{sec:accuracy}). This analysis shows that a resolution of 0.1'' is recommended to obtain a reliable measurement. 

\section{Discussion}\label{sec:discussion}

\subsection{Numerical resolution}
\label{sec:res}

\begin{figure}
    \centering
    \includegraphics[scale=0.43]{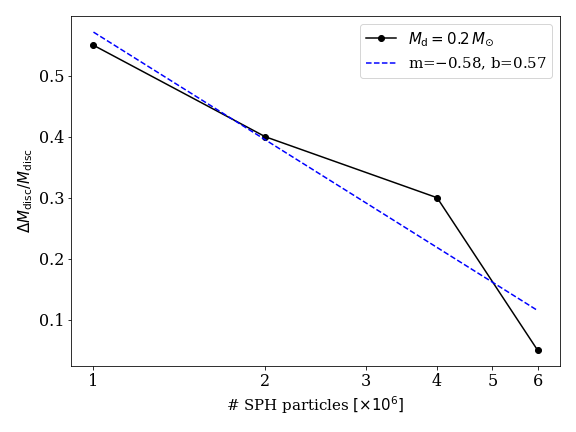}
    \caption{Uncertainties obtained from our fitting procedure for the disc mass $\Delta M_{\rm d}/M_{\rm d}$ for the combined fit procedure with different numerical resolutions, $[1,2,4,6] \times 10^6$ SPH particles. The blue dashed line is the semi-log linear fit of the data points with slope $m=-0.58$. As expected accuracy increases with numerical resolution.}
    \label{fig:res_num}
\end{figure}

Since the height of the emitting layer increases with the disc mass, in the \textbf{md0.2} simulation, we notice that the emitting layer is higher compared to the typical height of SPH particles, while this does not happen in the low disc mass case (for an example see Fig.~\ref{surface_mol}). Indeed, the number density of SPH particles decreases with $z$, since most of the disc mass is concentrated in the midplane. This implies that the size of the Voronoi grid cells used in \textsc{mcfost} is increasing with $z$, then resulting in artificial low resolution $^{12}$CO data-cubes for the high disc mass simulation. To test this, we perform additional hydrodynamical simulations with 2, 4 and 6 $\times 10^{6}$ SPH particles. Fig.~\ref{fig:res_num} shows that by increasing the number of SPH particles, the uncertainty we obtain for the disc mass decreases, until we reach the "true" disc mass value for the 6M particles simulation. The blue dashed line represents the semi-log linear fit of the data points.

\subsection{Limitations and further work}
The main assumptions adopted in this study are: 
\begin{itemize}
    \item[--] Model: axisymmetric discs, no magnetic field, no planets; 
    \item[--]Hydrodynamic simulations: gas only, since the dust-to-gas ratio is assumed to be so low as to not affect the result. The disc is assumed to be locally and vertically isothermal;
    \item[--] Radiative transfer: $T_{\rm gas} = T_{\rm dust}$ (LTE), Mie theory for optical properties, ISM-like grain size distribution, dust-to-gas ratio $\sim 0.01$.
    \item[--] Post-processing of synthetic images: the simulated ALMA images are obtained by convolving with a Gaussian beam and the noise has been added directly to the resulting images. Thus, the filtering of an interferometer and the noise properties of ALMA observations are not fully captured with these simple assumptions. As a test, we also apply the method presented in this work on a simulated ALMA image obtained with CASA, showing no difference in the fitted disc parameters.

\end{itemize}

Our study is restricted to stable self-gravitating discs, excluding those having peculiar substructures such as GI spirals. It would be important to reproduce the benchmark done in this study on more general cases, e.g. to confirm masses estimated for Elias 2-27, IM Lup and GM Aur. This study shows that one should investigate how the extraction of the rotation curve is affected by the presence of non-axisymmetric structures (e.g., by including cooling and allowing the disc to develop GI in the shape of spiral structures). Indeed, non-axysymmetric structures might be important in young objects which are still accreting from the molecular clouds (e.g., streamers). Other aspects that might be worth investigating consist in determining whether the presence of an embedded planet or a central binary could affect the disc mass estimate. 
Finally, the case of non-isothermal discs is going to be investigated by Martire et al. (in prep).

\section{Conclusions}\label{sec:conclusion}
In this study we have benchmarked the determination of disc masses using rotation curves extracted from channel maps as introduced in \cite{veronesi2021} and \cite{lodato23}. We generate controlled mock observations from \textsc{phantom} SPH simulations with different disc masses that are post-processed with the \textsc{mcfost} radiative transfer code. We adopt spatial and spectral resolution that are the same as the ones of the MAPS survey (0.1$''$ and 0.1 km/s).

We find that for $M_{\rm d}/M_{\star} \gtrsim 0.05$, we recover the correct disc-to-star mass ratio with a typical accuracy of $\sim 25 \%$, the uncertainties being equally spread between uncertainties over the aspect ratio and the inclination of the disc, as well as the height of the emitting layer (the truncation radius and the stellar mass are also obtained with uncertainties of $\sim 15\%$ and $\sim 8\%$ respectively). For lower resolutions (0.2-0.3''), disc masses recovered shift towards larger values ($0.1\,M_{\odot}$ for 0.2''). This constraint should be accounted for when proposing kinematics ALMA observations aiming at dynamically measuring the disc mass. This method can be used to calibrate already existing methods such as flux based methods.

\begin{acknowledgements}
The authors want to thank the referee for constructive comments that improved this manuscript. This work has received funding from the ERC CoG project PODCAST No 864965, and from the European Union’s Horizon 2020 research and innovation programme under the Marie Skłodowska-Curie grant agreement No 823823 (RISE DUSTBUSTERS project). S.F. is funded by the European Union (ERC, UNVEIL, 101076613). Views and opinions expressed are however those of the author(s) only and do not necessarily reflect those of the European Union or the European Research Council. Neither the European Union nor the granting authority can be held responsible for them.

We used the \textsc{emcee} algorithm \citet{emcee}, the \textsc{corner} package \citet{corner} to produce corner plots, and \textsc{python}-based MATPLOTLIB package \citet{matplotlib} for all the other figures. Computational resources were provided by the PSMN (Pôle Scientifique de Modélisation Numérique) of the ENS de Lyon, and the INDACO Platform, a project of High Performance Computing at the Università degli Studi di Milano. The authors thank Pietro Curone, Andrés F. Izquierdo, Sean Andrews, Richard Teague, Giovanni Rosotti and Enrico Ragusa for useful discussions.
\end{acknowledgements}

   \bibliographystyle{aa} 
   \bibliography{biblio} 

\newpage
\begin{appendix}

\section{Elias 2-27 updated fit results}\label{app:elias}
We derived with the updated model the disc mass of Elias 2-27 finding a result consistent with \cite{veronesi2021}. In particular, when fitting simultaneously the West and the East side of the disc, and the two isotopologues ($^{13}$CO and C$^{18}$O) we obtain a star mass $\sim0.40\,M_{\odot}$ and a disc mass $\sim0.13\,M_{\odot}$ ($\sim0.46\,M_{\odot}$ 
 and $\sim0.08\,M_{\odot}$ in \citealt{veronesi2021}).

\section{Dimensionless model}\label{app:resc}
Eq.~\ref{model} can be expressed in terms of $x=R/R_c$ and $\zeta =z/R_c$ as
\begin{equation}\label{dless_model}
    v_\phi^2 / \frac{GM_\star}{R_c} = f_1(x,\zeta) + \frac{1}{2 \pi} \frac{M_{\rm d}}{M_{\star}} f_2(x,\zeta),
\end{equation}
where
\begin{equation}
    f_1(x,\zeta) = \frac{1}{x} \left\{ 1- \left[ \gamma^\prime + (2-\gamma)x^{2-\gamma}\right]\left(\frac{H_{\rm c}}{R_{\rm c}}\right)^2 x^{1 - q} -q\left(1-\frac{1}{\sqrt{1+\zeta^2}}\right)\right\}, \nonumber
\end{equation}
and 
\begin{equation}
f_{2}(x,\zeta) = \int_{0}^{\infty} \Bigg[ K(y)-\frac{1}{4}\left(\frac{y^{2}}{1-y^{2}}\right) \times  \\ 
\left(\frac{u}{x}-\frac{x}{u}+\frac{\zeta^{2}}{x u}\right) E(y)\Bigg] \sqrt{\frac{u}{x}} y \tilde{\Sigma} \left(u\right) \mathrm{d} u\,. \nonumber
\end{equation}
where $H_{\rm c}$ denotes the scale height at the tapered radius $R_{\rm c}$, $y \equiv \frac{4u}{\left(u + x \right)^{2} + \zeta^{2}}$, $\tilde{\Sigma} \left(r / R_{\rm c}\right) \equiv \Sigma\left( r \right) / \frac{\tilde{M}_{\rm d}}{2 \pi R_{\rm c}^{2}}$ denotes the dimensionless surface density and $\tilde{M}_{\rm d} \equiv 2 \pi \int_{R_{\rm in}}^{R_{\rm out}} R \Sigma\left( R \right) \mathrm{d} R \simeq M_{\rm d}$ denotes the true mass of the disc. Hence, the azimuthal velocity can be rescaled in terms of disc to star mass ratio $M_d/M_\star$.

   \begin{table}
      \caption[]{Parameters of our SPH simulations. $M_\star$ is the star mass, and $M_{\rm disc}$ is the disc initial mass. $R_{\rm in}$ and $R_{\rm out}$ denote the initial disc inner and outer radii. The surface density profile of Eq.~\ref{eq:power-cutoff} is parametrised by the radius of the exponential tapering $R_{\rm c}$ and the power-law index $p$. $q$ is the power-law index of the temperature radial profile as in Eq.~\ref{eq:temperature}. $\alpha_{\rm ss}$ is the effective \citet{shakura73} viscosity. $(H/R)_{\rm c}$ is the disc aspect ratio at the reference radius $R_{\rm c}$. }
         \label{tab:setups_SPH}
     $$ 
         \begin{tabular}{p{0.35\linewidth}p{0.35\linewidth}}
            \hline
             Parameters &  Value \\ 
            \hline
            $M_\star\, [M_\odot]$ & $1$ \\ 
            $M_{\rm disc}\, [M_\odot]$ &$[0.01,0.025,0.05,$\\
            & $0.1,0.15,0.2]$ \\
            $R_{\rm in}\, [{\rm au}]$  & $10$ \\
            $R_{\rm out}\, [{\rm au}]$ & $300$ \\
            $R_{\rm c}\, [{\rm au}]$ & $100$ \\
            $\gamma$ & $1.$\\ 
            $q$ & $0.5$\\
            $\alpha_{\rm SS}$  & $ 0.005$    \\
            $(H/R)_{\rm c}$ & $0.075$ \\

            \hline
         \end{tabular}
     $$ 
   \end{table}

\section{Simulation parameters}
\label{app:params_simu}
The parameters used in the hydrodynamic simulations described in Sec.~\ref{sec:hydro} are given in Table~\ref{tab:setups_SPH}.

\section{Properties of the emitting layer}

\begin{figure}
    \centering
    \includegraphics[width=0.5\textwidth]{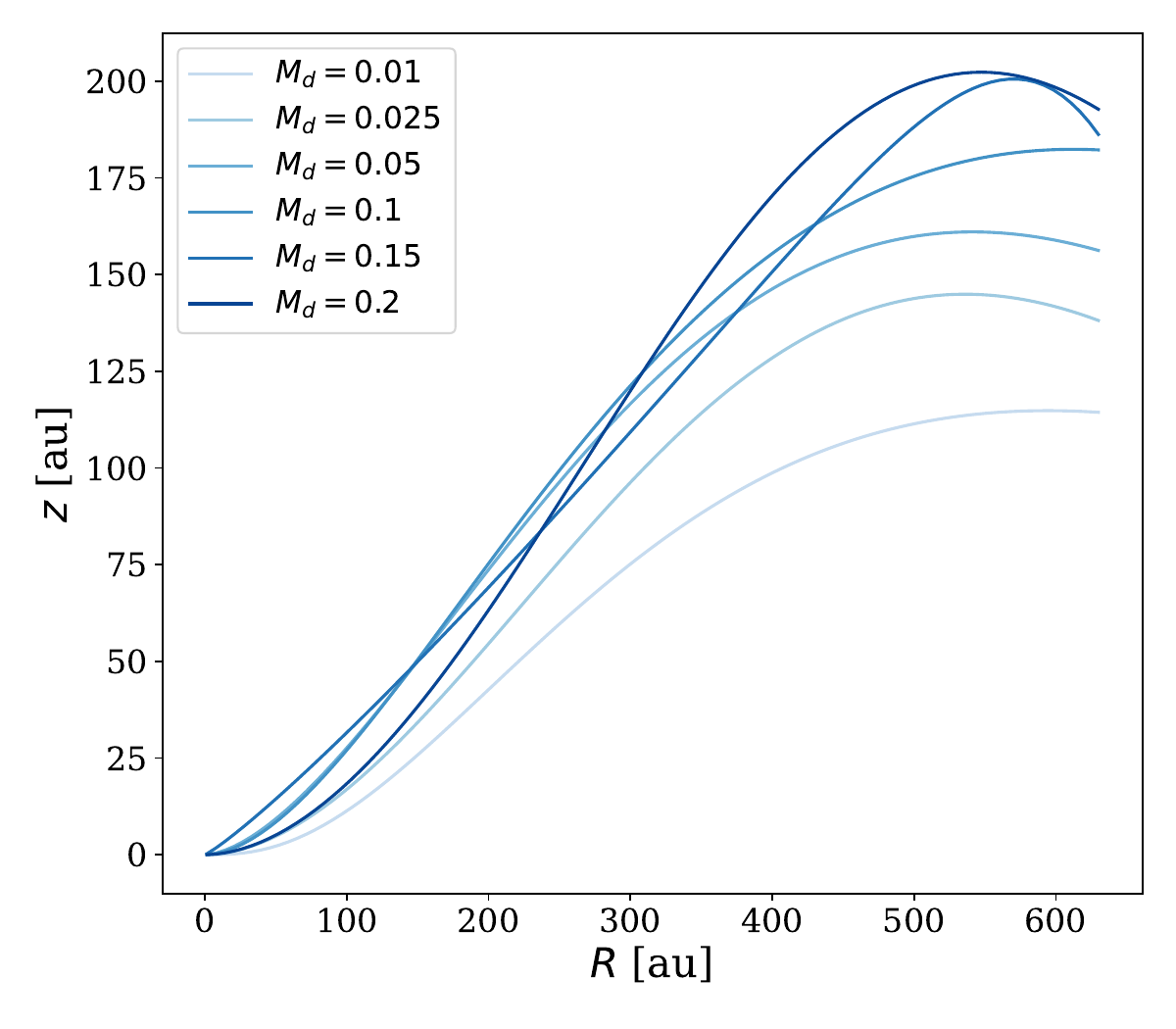}
    \caption{Emitting layer derived with \textsc{disksurf} for different disc masses for the $^{12}$CO isotopologue.}
    \label{surface_mass}
\end{figure}

\begin{figure}
    \centering
    \includegraphics[width=0.5\textwidth]{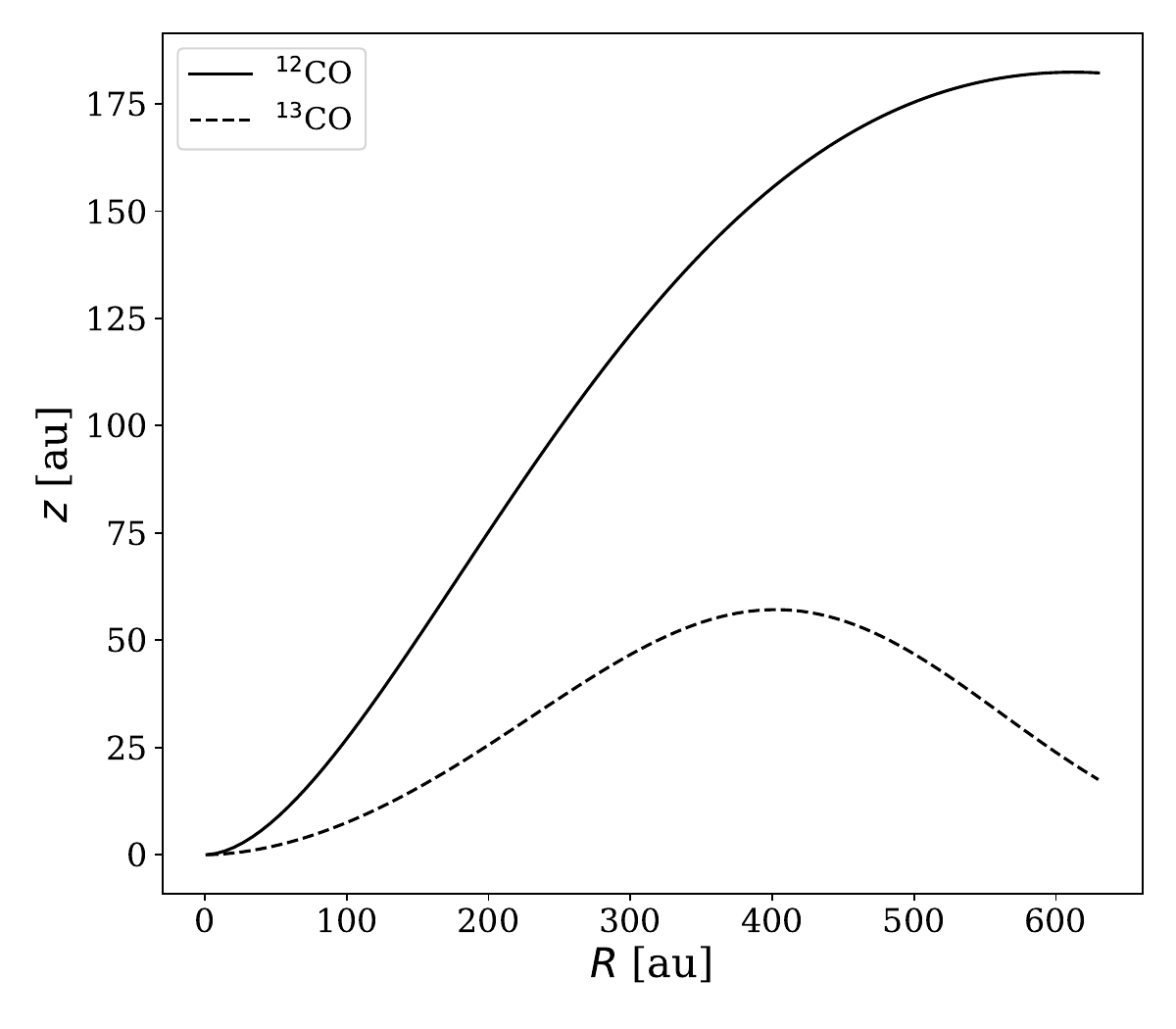}
    \caption{Comparison between the emitting layer of $^{12}$CO (solid line) and $^{13}$CO (dashed line) for a disc mass of 0.1 $M_{\odot}$.}
    \label{surface_13co_12co}
\end{figure}

Fig.~\ref{surface_mass} shows the height of emitting layers as a function of the radius for different disc masses. As expected, the emitting layer is higher for large disc masses, since the column density is larger. Fig.~\ref{surface_13co_12co} highlights the difference between the height of layers derived for the $^{12}$CO (orange line) and $^{13}$CO (blue line) isotopologues. $^{13}$CO is optically thinner compared to $^{12}$CO and traces therefore a region closer to the midplane, its emitting layer is also lower compared to the one of $^{12}$CO one. 

\begin{figure}
    \centering
    \includegraphics[width=0.5\textwidth]{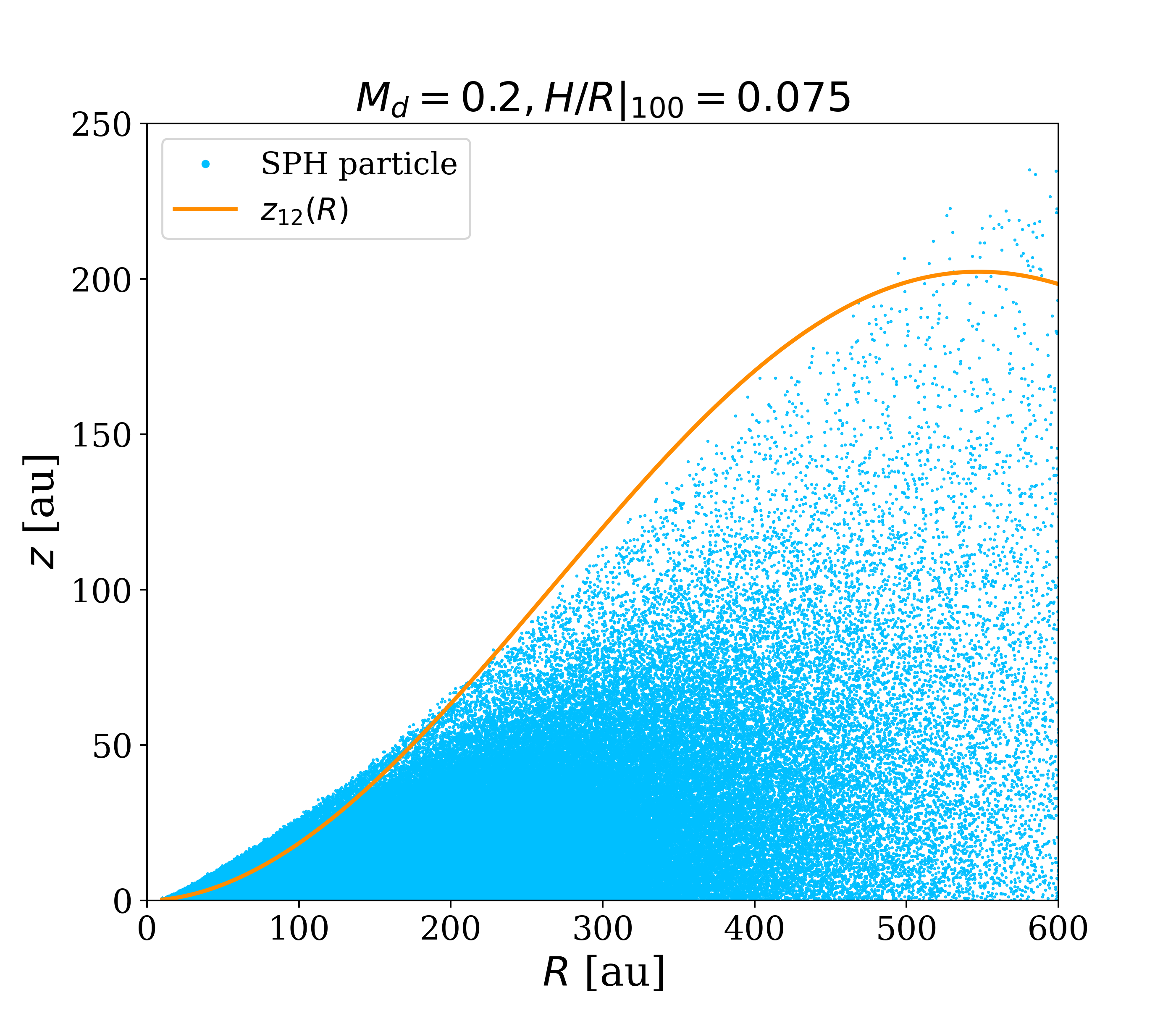}
    \includegraphics[width=0.5\textwidth]{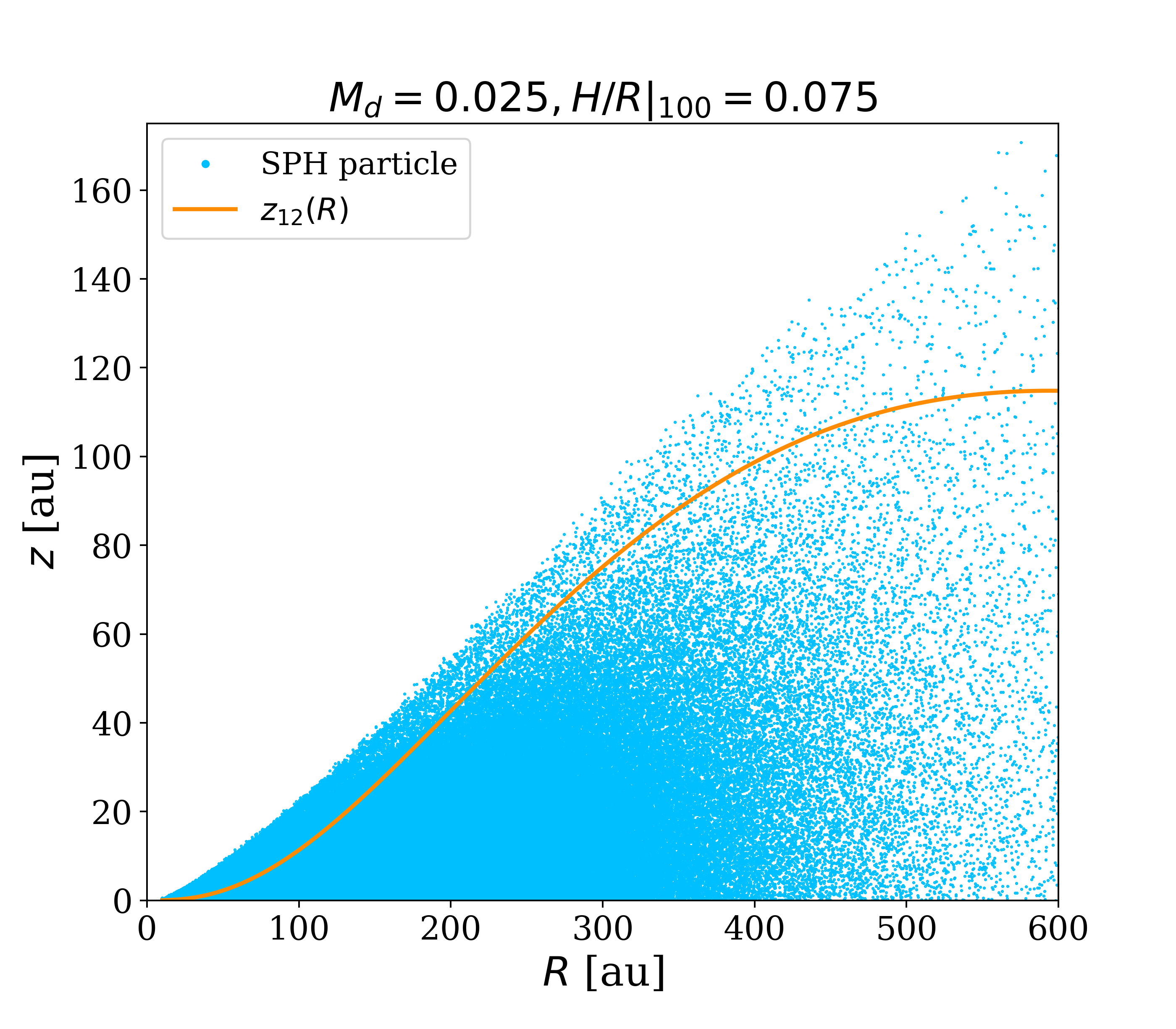}
    \caption{Distribution of SPH particles (cyan dots) and the emitting layer obtained with \textsc{disksurf} (orange line) for two simulations, one with high mass (upper panel) and one with low mass (lower panel).}
    \label{surface_mol}
\end{figure}

Fig.~\ref{surface_mol} shows the comparison between emitting layer derived for 0.2 $M_{\odot}$ (top panel) and 0.025 $M_{\odot}$ (bottom panel). Emitting layers are plotted with orange solid lines, while the SPH particles particles are represented with cyan markers. The number density of SPH particle decreases with $z$, since most of the disc mass tends to be concentrated in the midplane. This implies that the size of the Voronoi grid cells increase with $z$, resulting in an artificially low resolution in the upper layer of the $^{12}$CO data-cubes. This becomes problematic in the case of the high disc mass simulations (e.g. here 0.2 $M_{\odot}$ for $10^6$ SPH particles), since the disc is optically thicker and the emitting layer is higher with respect to the case of lighter discs. 

\section{Double Gaussian method}\label{app:doublegauss}

\begin{figure}
    \centering
    \includegraphics[scale=0.4]{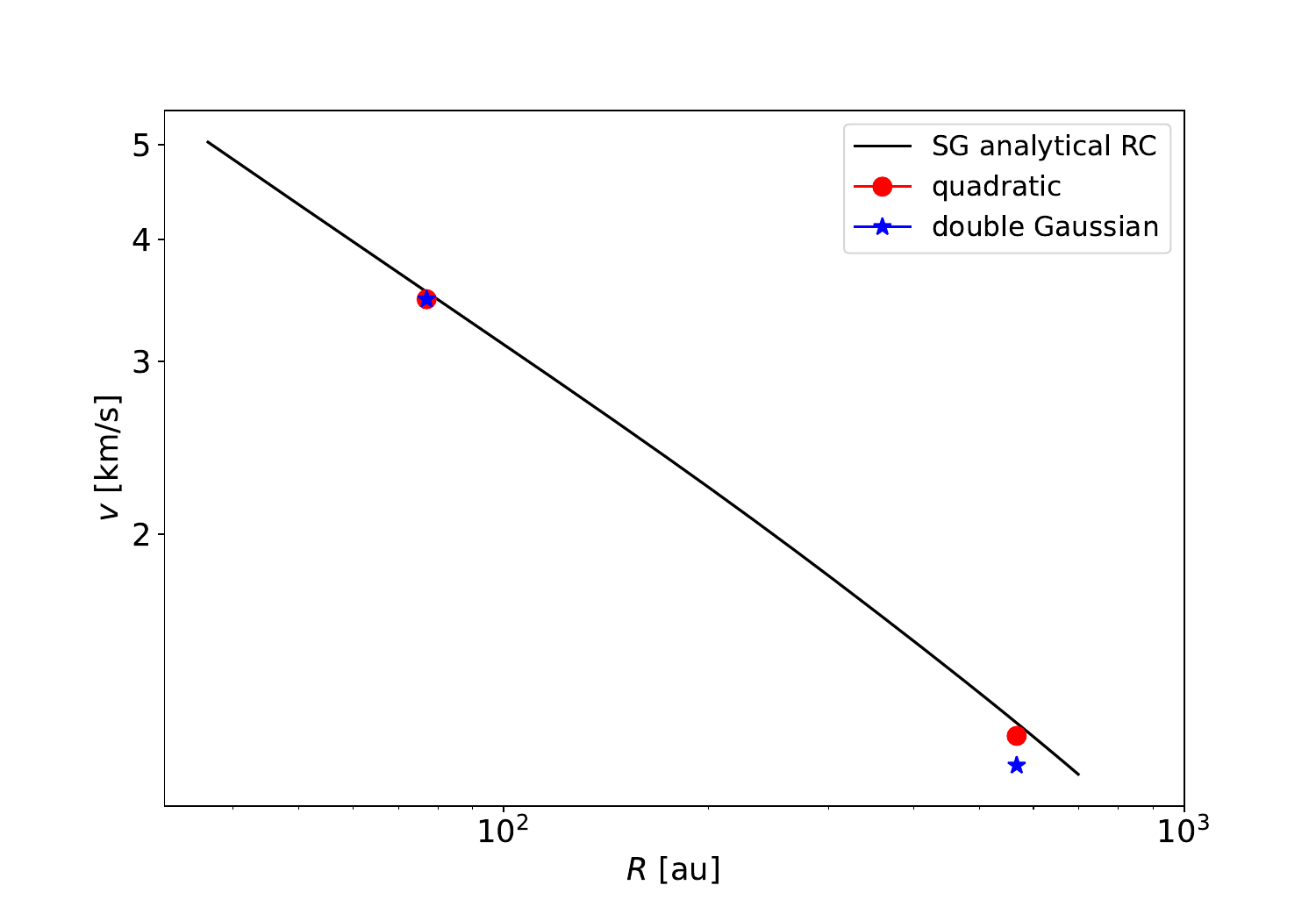}
    \caption{Comparison among the analytical rotation curve obtained with the self-graviting model (Eq.~\ref{model}, black solid line), and the velocity obtained with the quadratic (red circle) and the double Gaussian method (blue star), computed at two different radii (70 au and 560 au).}
    \label{fig:rotation_comparison}
\end{figure}

Fig.~\ref{fig:rotation_comparison} shows a comparison between the analytical rotation curve obtained with the self-gravitating model (Eq.~\ref{model}), and the velocity values obtained with the quadratic and the double Gaussian methods, computed at two different radii. We note that in the outer region the value extracted with the double Gaussian method deviates further from the expected value. In fact, in the outer region the S/N is too low, and the double Gaussian method fits the noise rather than the signal.

\section{Other uncertainties}\label{app:others}

\begin{figure*}
    \centering
    \includegraphics[scale=0.32]{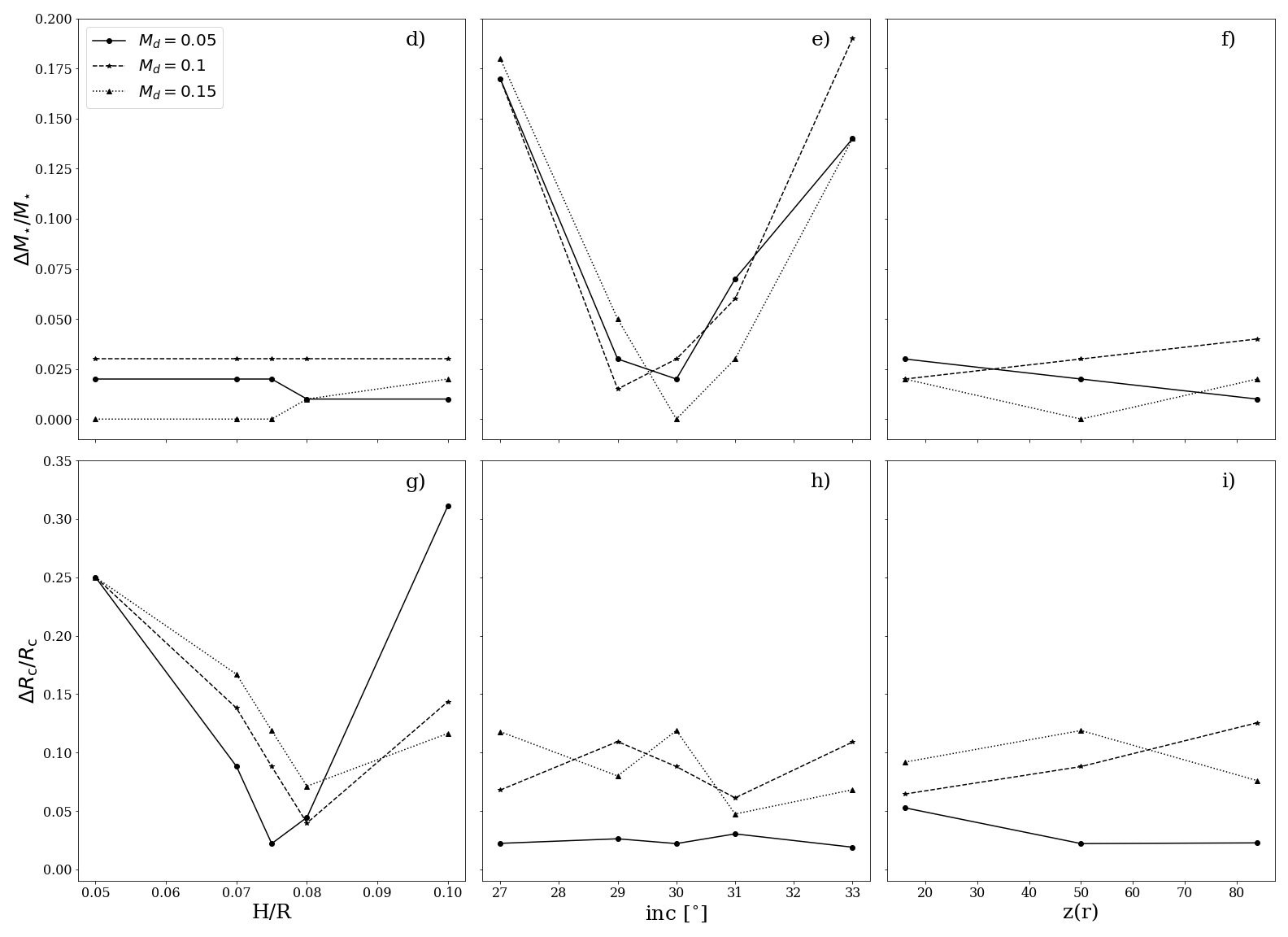}
    \caption{Uncertainties related to the fitting procedure for the mass of the star $M_{\star}$ (top row) and the disc truncation radius $R_{\rm c}$ (bottom row). Estimates are given as functions of the aspect ratio of the disc $H/R$ (left column), the disc inclination (middle column) and the disc emitting layer $z(R)$ (right column). Different markers and lines styles represent results obtained for simulations with different disc masses.}
    \label{fig:accuracy_star_rad}
\end{figure*}
Fig.~\ref{fig:accuracy_star_rad} shows the uncertainties for the mass of the star (top row) and the disc truncation radius (bottom row), as function of the aspect ratio of the disc $H/R$ (left), the disc inclination (centre) and the disc emitting layer $z(R)$ (right). While for the uncertainties for the disc mass (see Fig.~\ref{fig:accuracy_mass}), the three parameters $H/R$, $i$ and $z(r)$ have similar contributions
as sources of errors, now we can identify two main contributions. For the mass of the star and the disc truncation radius, the principal sources of uncertainties are the inclination of the disc (panel \textbf{e} in Fig.~~\ref{fig:accuracy_star_rad}) and its aspect ratio (panel \textbf{g} in Fig.~~\ref{fig:accuracy_star_rad}) respectively. Indeed, varying the inclination with respect to the expected value implies that the annulus along which the observed velocity is computed to extract the rotation curve are not at constant radius. Another source of error is related to the de-projection of the rotation curve. This affects strongly the fitted value of the mass of the star. As for the aspect ratio, it mostly influences the truncation radius, since it scales with $(H/R)^2$ (Eq. \ref{model}).

\end{appendix}
\end{document}